\documentclass[a4paper,11pt]{article}  

\usepackage{graphicx}
\newcommand{\lapprox}{\stackrel{<}{\sim}}

\begin{document}
\baselineskip 14pt

\pdfpagewidth 20.0cm
\pdfpageheight 11in

\begin{titlepage}

\title{{\bf Diagrammatic Pairing Fluctuations Approach to the BCS-BEC Crossover}}
\vspace{2cm}
\par
\author{G. C. Strinati \\ \emph{Dipartimento di Fisica, Universit\`{a} di Camerino} \\ \emph{I-62032 Camerino, Italy}}
\vspace{1cm}

\date{}
\maketitle
\vspace{1cm}

\begin{abstract}
This paper gives a survey of a diagrammatic approach for fermionic pairing fluctuations, which are relevant to the
BCS-BEC crossover realized with ultracold Fermi gases.
Emphasis will be given to the physical intuition about the relevant physical processes that can be associated with this approach.
Specific results will be presented for thermodynamic and dynamical quantities, where a critical comparison with alternative
diagrammatic approaches will also be attempted.
\end{abstract}

\vspace{2.5cm}

\begin{center}
{\bf -----------------------------------------------------------------------------------}
\end{center}

\noindent
[To be published as a chapter in {\bf BCS-BEC Crossover and the Unitary Fermi Gas} 
(Lecture Notes in Physics), edited by Wilhelm Zwerger (Springer, 2011).]

\thispagestyle{empty}
\end{titlepage}

\newpage
\section{Introduction}
\label{sec:introduction}

The BCS-BEC crossover has been of considerably interest over the last several years, especially after its experimental realization
with ultracold Fermi ($^{6}\mathrm{Li}$ and $^{40}\mathrm{K}$) gases \cite{reviews-2008}.
By this approach, a \emph{continuous evolution} is sought from a BCS-like situation whereby Cooper pairs are highly overlapping, to a BEC-like situation where composite bosons form out of fermion pairs and condense at sufficiently low temperature.
Here, reference to composite bosons stems from the fact that the temperatures of formation and condensation are in this case comparable with each other, in contrast with more conventional point-like bosons for which the two temperatures are quite different (reflecting the fact that their internal structure has no relevance to problems related to condensation). 
Accordingly, a theoretical description of composite bosons should take into account not only their overall bosonic structure associated
with the center-of-mass motion, but also their composite nature in terms of the degrees of freedom of the constituent fermions.

\begin{figure}[htbp]
\begin{center}
\includegraphics[width=8.5cm,angle=0]{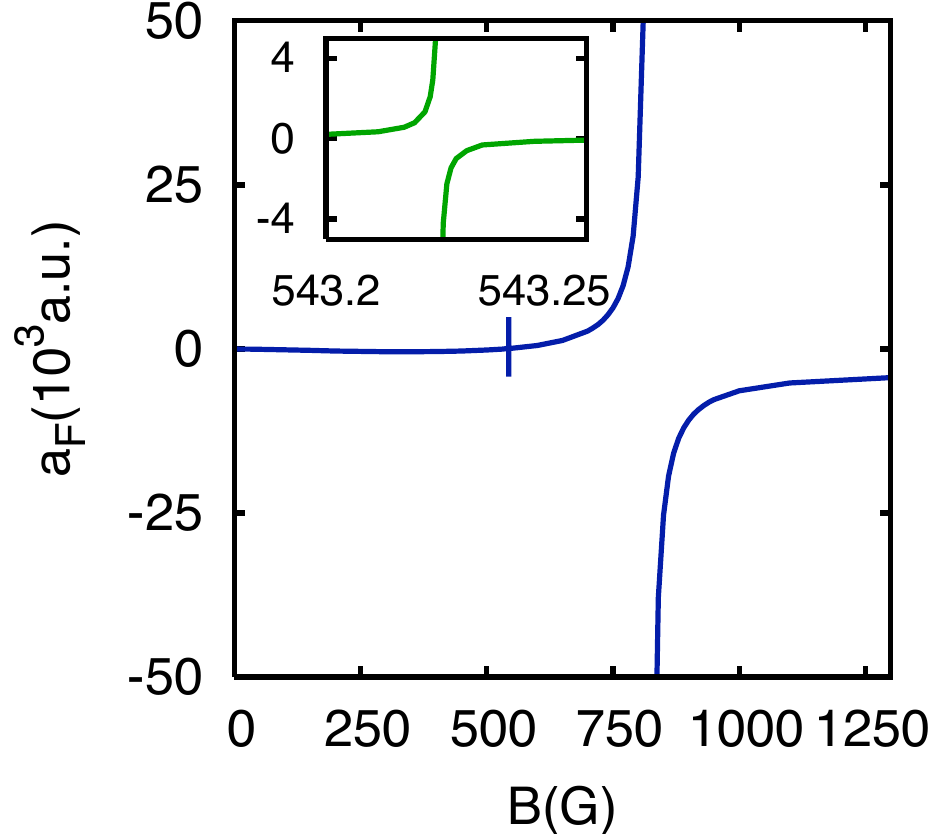}
\caption{Scattering length for $^{6}\mathrm{Li}$ atoms vs magnetic field. The inset amplifies the behavior of the narrow resonance.
             (Adapted from Fig.2 of Ref.\cite{SPS-2005}.)}
\label{figure0}
\end{center}
\end{figure}

The key feature of ultracold Fermi atoms that has allowed the realization of the BCS-BEC crossover is the possibility of varying essentially at will the strength of the attractive interaction between fermions of different species \cite{FF-resonances-e}, attraction which results in the formation of Cooper pairs in a medium, on the one hand, and of composite bosons in vacuum, on the other hand, out of the two fermion species.
[In the case of ultracold atoms, the spin of an electron is replaced by an analogous quantum number associated with the atomic hyperfine levels.]
Owing to this unique possibility, ultracold atoms should be regarded as prototype systems, with respect to others in Nature for which this possibility is hindered.
Specifically, in ultracold atomic gases the attractive interaction is varied through the use of the so-called \emph{Fano-Feshbach resonances}, which are characterized by a resonant coupling between the scattering state of two atoms with near-zero energy and a bound state in a close channel \cite{FF-resonances-t}.
Changing (through the variation of a static magnetic field $B$) the position of the bound state with respect to threshold in a suitable way, one can modify the value of the (fermionic) \emph{scattering length} $a_{F}$ from negative values before the formation of 
the bound state in the two-body problem to positive values once the bound state is formed \cite{Fano-Rau}.
As an example, Fig.\ref{figure0} shows the scattering length for the collision of two $^{6}\mathrm{Li}$ atoms vs $B$. 

In this context, the dimensionless parameter $(k_{F}a_{F})^{-1}$ acquires a special role for the corresponding many-body system 
at finite density. 
Here, the Fermi wave vector $k_{F}$ of the non-interacting system (which is defined as $k_{F} = \sqrt{2 m E_{F}}$
both for a homogeneous and a trapped system - see below) is a measure of the (inverse of the) interparticle distance, $m$ being the fermion mass and $E_{F}$ the Fermi energy of non-interacting fermions (we set $\hbar = 1$ throughout).
When the Fano-Feshbach resonance is sufficiently ``broad'' (like for $^{6}\mathrm{Li}$ and $^{40}\mathrm{K}$ atoms used
in experiments thus far), in fact, the many-body fermion problem can be described in a simplified way by a single-channel Hamiltonian with an instantaneous short-range interaction \cite{SPS-2005}.
The strength of this interaction, in turn, can be parametrized in terms of the above scattering length $a_{F}$ of the two-body (molecular) problem, which shares the same ultraviolet divergency associated with the short-range character of the two-body potential \cite{Randeria-93,PS-00}.
In this way one ends up with all physical quantities of interest for the many-body system depending on the interaction only 
through the parameter $(k_{F}a_{F})^{-1}$.

In terms of this parameter, one finds that for most physical quantities the crossover between the BCS and BEC regimes is exhausted, 
in practice, within a range $\approx 1$ about the unitary limit at $(k_{F}a_{F})^{-1} = 0$ where $a_{F}$ diverges.
Outside this limited range, the BCS and BEC regimes (whereby $(k_{F}a_{F})^{-1} \lapprox -1$ and 
$1 \lapprox (k_{F}a_{F})^{-1}$, in the order) are characterized by the product $k_{F}|a_{F}|$ being quite smaller than unity
(corresponding to a diluteness condition), so that theoretical approaches can in principle be controlled in terms of this small quantity
in these two separate regimes.
No such small parameter evidently exists, however, in the unitary regime about $(k_{F}a_{F})^{-1}=0$, whose theoretical description consequently constitutes a formidable task. 

It is then clear that theoretical treatments of the BCS-BEC crossover should provide as accurate as possible descriptions 
of the two regimes where the above diluteness condition applies, either in terms of the constituent fermions (BCS regime) or
of the composite bosons (BEC regime).
Specifically, this has to occur via a \emph{single} fermionic theory that bridges across these two limiting representations, by
recovering controlled approximations on both sides of the crossover and providing at the same time a continuous evolution between them, thereby spanning also the unitary regime where use of the theory could \emph{a priori} not be justified.

The prototype of this kind of approach is represented by the BCS theory itself at \emph{zero temperature}.
As remarked originally by Leggett \cite{Leggett} (see also Ref.\cite{Eagles}), the BCS wave function is quite more 
general than originally thought, in the sense that it contains as an appropriate limit the coherent state associated with a Bose-Einstein condensate of composite bosons made up of opposite-spin fermions.
This limit is reached when the occupation numbers of all possible fermionic single-particle states are much less than unity, so that the Fermi surface is completely washed out.

\begin{figure}[htbp]
\begin{center}
\includegraphics[width=12.5cm,angle=0]{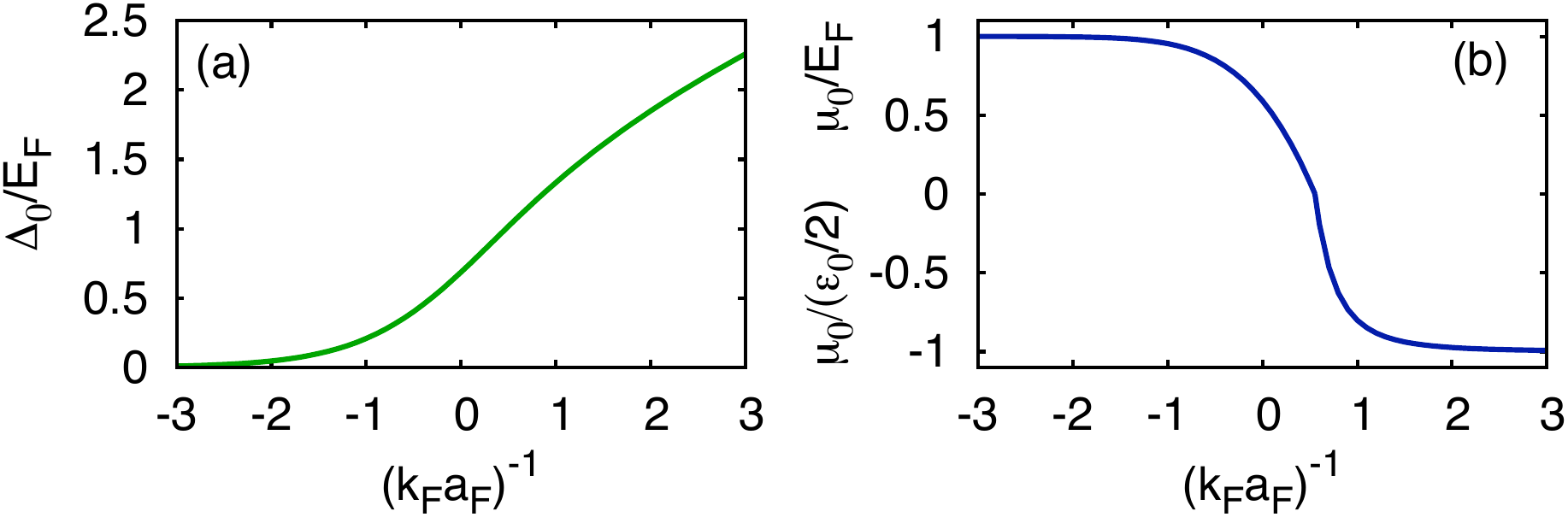}
\caption{(a) Gap parameter $\Delta_{0}$ and (b) chemical potential $\mu_{0}$ for a homogeneous system at zero temperature 
              vs the coupling parameter $(k_{F} a_{F})^{-1}$, evaluated across the BCS-BEC crossover within mean field.}
\label{figure1}
\end{center}
\end{figure}

The argument can be made more quantitative by solving the coupled gap and density equations provided by the BCS theory \cite{BCS,Schrieffer} for a homogeneous system at $T=0$ (for which an analytic solution exists in terms of elliptic integrals \cite{MPS}), 
to obtain the gap (order) parameter $\Delta_{0}$ and fermionic chemical potential $\mu_{0}$ as functions of $(k_{F}a_{F})^{-1}$ as shown in Fig.\ref{figure1}.
Note that the chemical potential crosses over as expected, from the value $E_{F}$ of the Fermi energy of non-interacting fermions in the BCS limit, to (half the value of) the binding energy $\varepsilon_{0} = (m a_{F}^{2})^{-1}$ of the two-body (molecular) problem within the single-channel model in the BEC limit. 
In both limits, $\Delta_{0}/|\mu_{0}| \ll 1$ albeit for different physical reasons.

\begin{figure}[htbp]
\begin{center}
\includegraphics[width=9.0cm,angle=0]{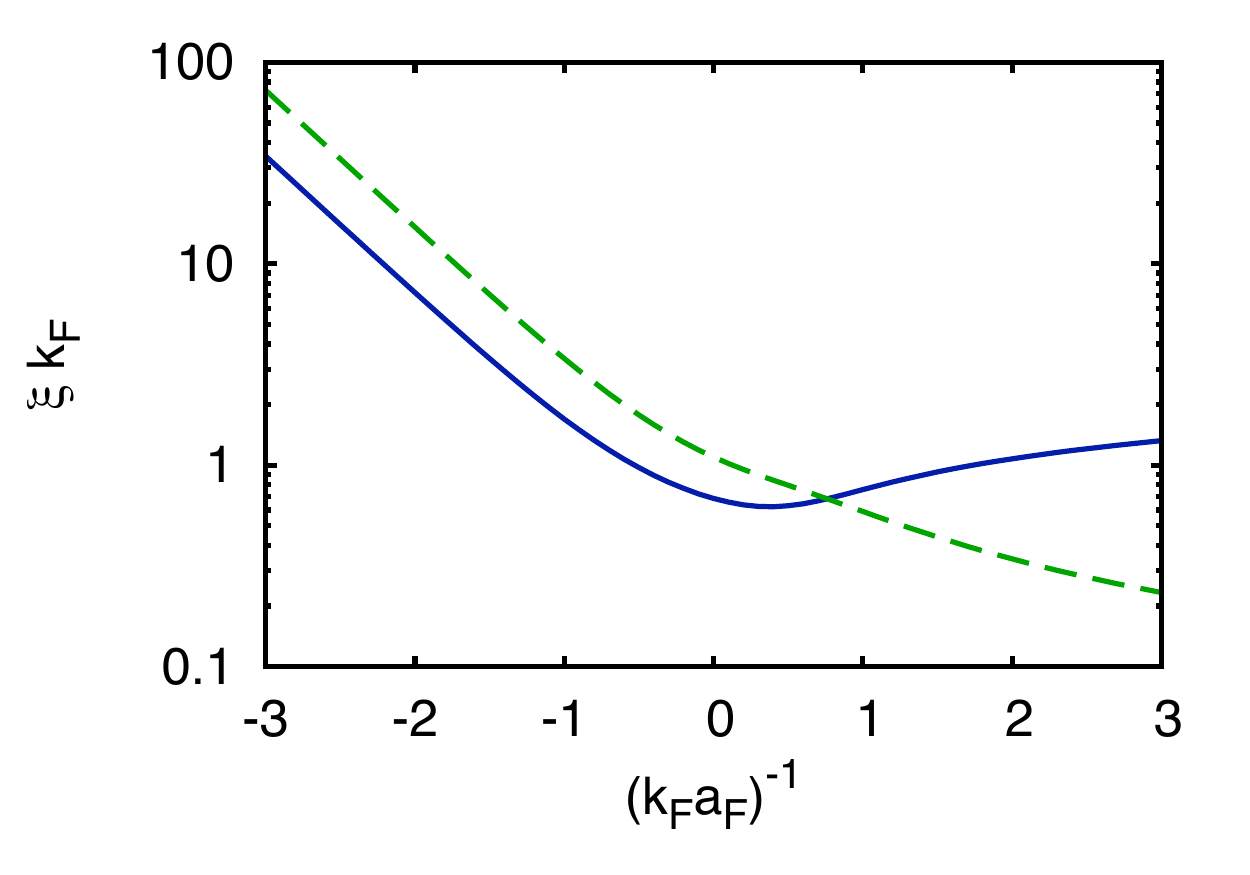}
\caption{Pair coherence length $\xi_{\mathrm{pair}}$ (dashed line) and phase coherence (healing) lenght $\xi_{\mathrm{phase}}$ 
             (full line) vs $(k_{F} a_{F})^{-1}$, evaluated at zero temperature as in Ref.\cite{MPS} according to their definitions given 
             in Refs.\cite{PS-94} and \cite{PS-96}, respectively.}
\label{figure2}
\end{center}
\end{figure}

The BCS theory is a mean-field approximation which relies on the Cooper pairs being highly overlapping in real space \cite{BCS}, 
so that their effects can be dealt with ``on the average''. 
As such, it is expected to be a valid approximation even \emph{at finite temperature} whenever this condition is satisfied.
It should accordingly apply to the BCS limit of the BCS-BEC crossover, but not to the unitary or BEC regimes where the typical
length scale for correlation between two fermions with different spins becomes comparable with the interparticle spacing $k_{F}^{-1}$.
This is shown in Fig.\ref{figure2} where the (zero-temperature) \emph{intra-pair} coherence length $\xi_{\mathrm{pair}}$ is plotted 
vs $(k_{F}a_{F})^{-1}$.

It is then evident that, away from the BCS limit, inclusion of \emph{fluctuation corrections} beyond mean field becomes essential to 
account for the relevant physical properties of the system.
An equivalent way of stating the problem is that, away from the BCS limit, the intra- and \emph{inter-pair} coherence lengths 
are expected to differ considerably from each other.
This is also shown in Fig.\ref{figure2} where the inter-pair coherence length $\xi_{\mathrm{phase}}$ is reported for comparison.
[In the BCS limit the two lenghts differ by an irrelevant numerical factor ($\xi_{\mathrm{pair}} \simeq (3/\sqrt{2}) \xi_{\mathrm{phase}}$) owing to their independent definitions, so that the two curves in Fig.\ref{figure2} are parallel to each other in this coupling regime.]
In particular, in the BEC regime $\xi_{\mathrm{pair}}$ corresponds to the size of a composite boson while $\xi_{\mathrm{phase}}$ represents the healing length associated with spatial fluctuations of the center-of-mass wave function of composite bosons.

It was indeed within this framework that the BCS-BEC crossover attracted attention also for high-temperature (cuprate) superconductors \cite{Uemura-1,Uemura-2}, for which the product $k_{F} \xi_{\mathrm{pair}}$ was estimated to be about 5$\div$10 in contrast with more conventional superconductors for which it is of the order $10^{3}\div10^{5}$.
Several theoretical works were then put forward on the BCS-BEC crossover 
in this context \cite{Randeria-90,Haussmann-93,PS-94,Zwerger-97,Levin-97-98}, with the limitations, however, that the origin and characteristics of the attractive interaction at the basis of this crossover were not known for cuprate superconductors.
These limitations have eventually been fully removed with the advent of ultracold Fermi atoms, to which we shall limit our considerations in the following.

One related reason to invoke the inclusion of fluctuation corrections beyond mean field stems from the values obtained within BCS theory for the critical temperature at which the order parameter vanishes.
Only a numerical solution of the coupled gap and density equations is amenable for generic values of $(k_{F} a_{F})^{-1}$, but analytic results can still be obtained in the BCS and BEC limits. 
One gets \cite{Randeria-93}:
\begin{equation}
k_{B} T_{c} \, \simeq \, \frac{8 \, E_{F} \, e^{\gamma}}{\pi \, e^{2}} \, 
                         \exp\{\pi/(2a_{F}k_{F})\}                                           \label{T-c-weak-coupling}
\end{equation} 

\noindent
in the BCS limit (where $k_{B}$ is Boltzmann constant and $\gamma$ Euler constant with $e^{\gamma}/\pi \simeq 0.567$), and
\begin{equation}
k_{B} T_{c} \, \simeq \, \frac{\varepsilon_{0}}
{2 \, \ln \left( \varepsilon_{0}/E_{F} \right)^{3/2}}                                    \label{T-c-strong-coupling}
\end{equation}

\noindent
in the BEC limit, respectively.
While the result (\ref{T-c-weak-coupling}) corresponds to what is familiar from BCS theory for weak coupling \cite{BCS}, 
the result (\ref{T-c-strong-coupling}) does \emph{not} coincide with what one would expect in the BEC limit, namely, the expression of the Bose-Einstein condensation temperature $k_{B} T_{\mathrm{BEC}} \, = \, 3.31 \, n_{B}^{2/3}/(2m)$ where $n_{B} = n/2$ is the density of composite bosons in terms of the density $n$ of the constituent fermions.
On the contrary, the expression (\ref{T-c-strong-coupling}) increases without bound when approaching the BEC limit for
$1 << (k_{F} a_{F})^{-1}$.

\begin{figure}[htbp]
\begin{center}
\includegraphics[width=10.5cm,angle=0]{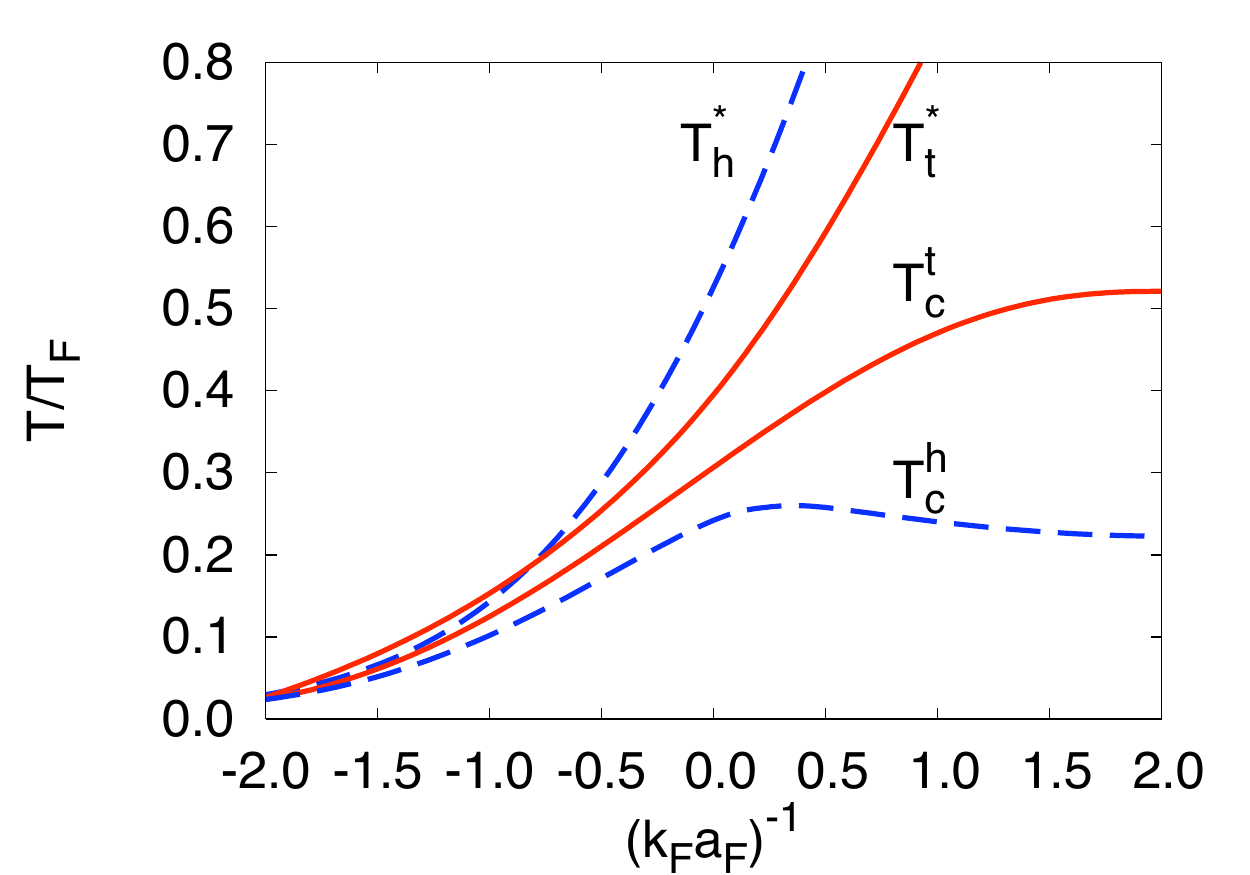}
\caption{Temperature vs coupling diagram for the trapped (full lines) and homogeneous (dashed lines) system,
             where the critical temperature $T_{c}$ and pair-breaking temperature $T^{*}$ are shown. 
             Each temperature is normalized to the respective Fermi temperature $T_{F}$. 
             (Adapted from Fig.1 of Ref.\cite{PPPS-04}.)}
\label{figure3}
\end{center}
\end{figure}

The points is that the critical temperature obtained from the solution of the mean-field equations corresponds to the process
of pair formation and not of pair condensation.
The two temperatures coincide only in the BCS (weak-coupling) limit \cite{GMB-61}, where pairs form and condense at the same time.
In the BEC (strong-coupling) limit, on the other hand, pairs form at a higher temperature than that at which they eventually condense owing to quantum effects.
Accordingly, the expression (\ref{T-c-strong-coupling}) signals the phenomenon of pair dissociation, and as such it must be regarded as a ``crossover'' temperature $T^{*}$ which does not correspond to a true phase transition.
The complete plot of $T^{*}$ obtained by solving numerically the mean-field gap and density equations throughout the BCS-BEC crossover is shown for the homogeneous (h) and trapped (t) cases in Fig.\ref{figure3}, where it corresponds to the upper dashed and full lines, respectively (the remaining two curves labeled by $T_{c}$ result instead beyond mean field and will be discussed in Section~\ref{sec:pairing-fluctuations}).

It is thus evident from the above discussion that the main limitation of the mean-field description we have considered thus far is that it includes only the degrees of freedom internal to the pairs which are associated with pair-breaking, but omits completely the translational ones.
The latter are responsible for the collective sound mode, which represents the main source of elementary excitations in the 
BEC regime \cite{SPS-09}.
To overcome this severe limitation for a sensible description of the BCS-BEC crossover in terms of a fermionic theory, it is then necessary to go beyond mean field and include \emph{pair-fluctuation} effects as discussed in the next Section.

\section{Inclusion of pairing fluctuations}
\label{sec:pairing-fluctuations}

\begin{figure}[htbp]
\begin{center}
\includegraphics[width=11.0cm,angle=0]{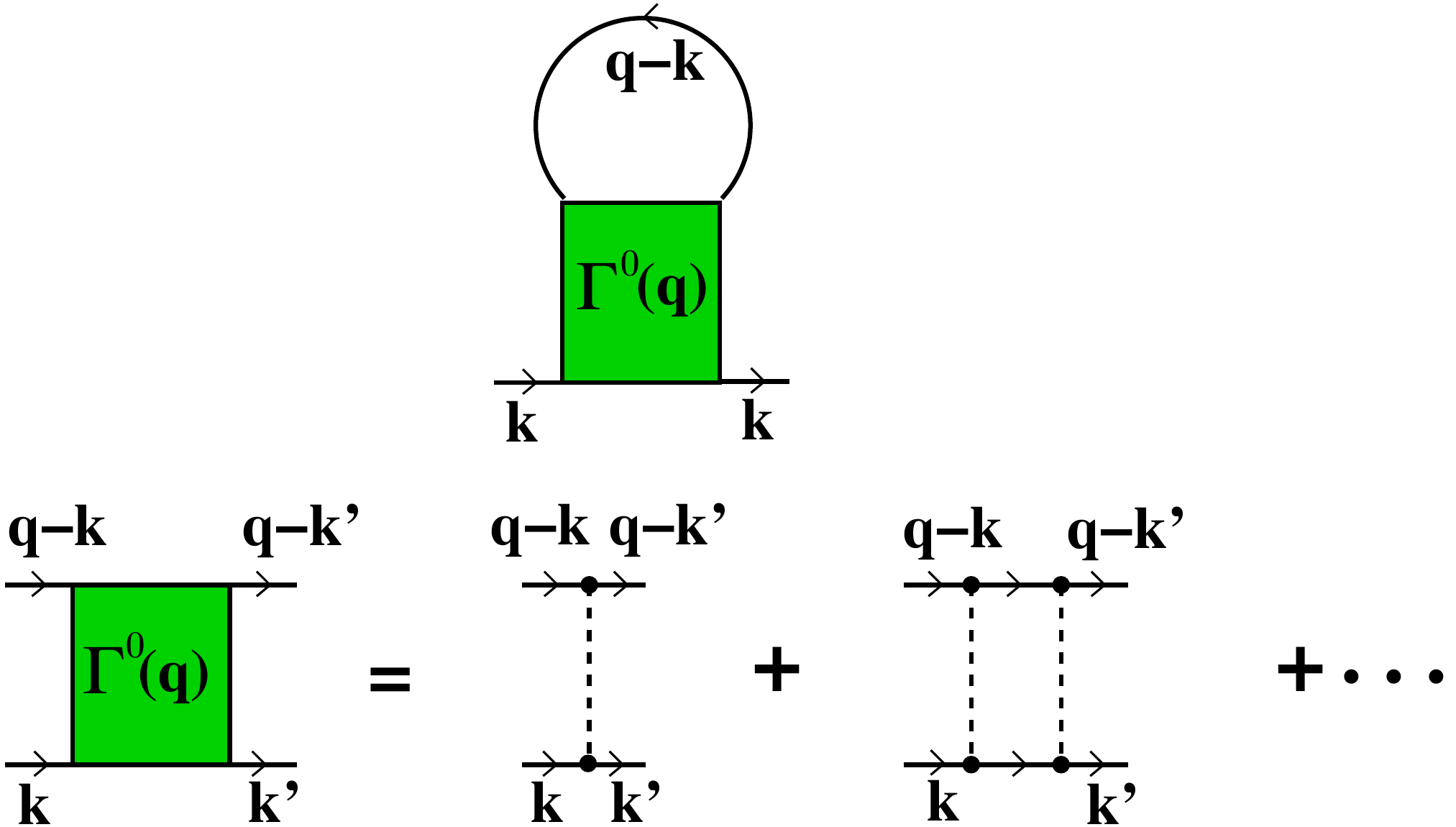}
\caption{Single-particle fermionic self-energy in the normal phase (upper panel) expressed in terms of the pair (ladder) 
              propagator $\Gamma^{0}$ between two fermions of opposite spins (lower panel).
              Full and dashed lines represent the fermionic propagator and interaction potential, respectively, while the
              labels $k$ ($k'$) and $q$ correspond to fermionic and bosonic four-vectors, in the order.}
\label{figure4}
\end{center}
\end{figure}

A diagrammatic approach for fermionic pairing fluctuations was first considered by Galitskii \cite{Galitskii} for a 
dilute Fermi gas with strong short-range repulsion \cite{FW}.
There it was shown that the relevant fermionic self-energy can be taken of the form depicted in Fig.\ref{figure4}, where 
$\Gamma^{0}$ is the pair propagator describing the repeated scattering in the medium between two fermions of opposite spins.

The short-range nature of the potential requires one to introduce at the outset a regularization procedure that eliminates the ultraviolet divergences.
This is done by exploiting the two-fermion problem in vacuum, which shares the same sort of divergences and for which the 
(positive) strength $v_{0}$ of the repulsive interaction can be related to the ultraviolet cutoff $k_{0}$ in wave-vector space through the following equation for the two-body t-matrix in the low-energy limit \cite{RT}:
\begin{equation}
\frac{m}{4 \pi a_{F}} \, = \, \frac{1}{v_{0}} \, + \, \int^{k_{0}} \! \frac{d\mathbf{k}}{(2\pi)^{3}} \, \frac{m}{\mathbf{k}^{2}}  \,\, .    
                                                                                                                                               \label{aF-vs-v0}
\end{equation}
This relation defines the (fermionic) scattering length $a_{F}$, which is positive in this case and remains smaller than the range
$\pi/(2 k_{0})$ of the potential if its strength $v_{0}$ is kept finite.

To the leading order in $a_{F}$, this self-energy results in a repulsive ``mean-field shift'' $(4 \pi a_{F}/m) \, n/2$ of the chemical potential, where $n$ is the total fermion density for both spin components.
This is because, to the leading order in $a_{F}$, $\Gamma^{0}(q) \simeq (-4 \pi a_{F}/m)$ and the loop in the upper panel of 
Fig.\ref{figure4} gives half the fermionic density $n$ (we consider throughout  the case of equal populations of spin up and  down fermions).
Terms up to the second order in $a_{F}$ were also obtained by the Galitskii original approach \cite{Galitskii}.

The above choice of the self-energy emphasizes \emph{pair-fluctuation effects} via the repeated scattering of two (opposite spin) 
fermions in the medium. 
As such, it has been considered physically relevant \emph{also} to the case of an attractive short-range potential with a negative $v_{0}$ \cite{Randeria-90,PPSC-02,Combescot-03}, for which $a_{F}$ has the typical resonant behavior of 
Fig.\ref{figure0} associated with the BCS-BEC crossover.
By this extrapolation, the formal structure of the Galitskii self-energy is carried over to the domain of strong coupling, and even further to the repulsive side of the resonance where a molecular state forms \cite{VSS-07}.

Let's write down explicitly the analytic expressions corresponding to the diagrams depicted in Fig.\ref{figure4}, for the simplest case 
when all fermionic propagators appearing therein are ``bare'' ones \cite{PPSC-02}.
One has:
\begin{equation}
\Sigma (\mathbf{k},\omega_{n}) = -k_{B} T \sum_{\nu} \int \! \frac{d\mathbf{q}}{(2\pi)^{3}} \,
\Gamma^{0}(\mathbf{q},\Omega_{\nu}) \, G_{0}(\mathbf{q}-\mathbf{k},\Omega_{\nu}-\omega_{n})
                                                                                                                                                              \label{pairing-self-energy}
\end{equation}

\noindent
for the fermionic self-energy, and
\begin{equation}
\frac{(-1)}{\Gamma^{0}(\mathbf{q},\Omega_{\nu})} = \frac{m}{4\pi a_{F}} + \int \! \frac{d\mathbf{k}}{(2\pi)^{3}}
\left[ k_{B} T \sum_{n} G_{0}(\mathbf{k},\omega_{n}) G_{0}(\mathbf{q}-\mathbf{k},\Omega_{\nu}-\omega_{n}) - \frac{m}{|\mathbf{k}|^{2}} \right]
                                                                                                                                                             \label{pairing-propagator}
\end{equation}

\noindent
for the (inverse of the) pair propagator.
Here, $G_{0}(\mathbf{k},\omega_{n}) = [i\omega_{n} - \xi (\mathbf{k})]^{-1}$ is the bare fermion propagator 
($\xi (\mathbf{k}) = \mathbf{k}^{2}/(2m) - \mu$ being the free-particle dispersion measured with respect to the chemical potential 
$\mu$), while $\omega_{n} = \pi k_{B}T(2n+1)$ ($n$ integer) and $\Omega_{\nu} = 2 \pi k_{B}T \nu$ ($\nu$ integer) are fermionic and bosonic Matsubara frequencies at temperature $T$, in the order.
Note how the strength $v_{0}$ of the attractive interparticle potential has been eliminated in the expression (\ref{pairing-propagator}) in favor of the scattering length $a_{F}$ via the relation (\ref{aF-vs-v0}), which now admits also negative value for $a_{F}$ consistently with the behavior shown in Fig.\ref{figure0}.

With the self-energy (\ref{pairing-self-energy}) one dresses the bare fermion propagator to obtain the full propagator
\begin{equation}
G(\mathbf{k},\omega_{n}) \, = \, \frac{1}{G_{0}(\mathbf{k},\omega_{n})^{-1} \, - \, \Sigma (\mathbf{k},\omega_{n})} \,\, ,
                                                                                                                                                  \label{dressed-fermion-propagator}
\end{equation}

\noindent
in terms of which the chemical potential can be eventually eliminated in favor of the density via the expression ($\eta = 0^{+}$):
\begin{equation}
n \, = \, 2 \, k_{B} T \sum_{n} e^{i \omega_{n} \eta} \int \, \frac{d\mathbf{k}}{(2\pi)^{3}} \, G(\mathbf{k},\omega_{n}) \,\, .
                                                                                                                                                  \label{fermion-density}
\end{equation}

On physical grounds, the relevance of the expressions (\ref{pairing-self-energy}) and (\ref{pairing-propagator}) to the 
BCS-BEC crossover can be appreciated from the following considerations.
While in the BCS weak-coupling limit (where $a_{F}<0$ and $(k_{F}a_{F})^{-1} \ll -1$) the pair propagator maintains formally the same expression $\Gamma^{0}(q) \simeq -4 \pi a_{F}/m$ of the repulsive case, in the BEC strong-coupling limit (where $0< a_{F}$ 
and $1 \ll (k_{F}a_{F})^{-1}$) it acquires the polar structure of a free-boson propagator \cite{Haussmann-93,PS-00}:
\begin{equation}
\Gamma^{0}(\mathbf{q},\Omega_{\nu}) \, = \, - \frac{8\pi}{m^{2} a_{F}} \, \frac{1}{i\Omega_{\nu} \, - \, 
\mathbf{q}^{2}/(4m) \, + \, \mu_{B}}                                                                                                \label{bosonic-propagator}
\end{equation}

\noindent
where the bosonic chemical potential $\mu_{B}$ reduces to $2\mu+\varepsilon_{0}$ in this limit when
the composite bosons have size $\approx a_{F}$.
In this limit, we may expand the fermionic propagator (\ref{dressed-fermion-propagator}) to the lowest order in $\Sigma$
\begin{equation}
G(\mathbf{k},\omega_{n}) \simeq G_{0}(\mathbf{k},\omega_{n}) \, + \, 
G_{0}(\mathbf{k},\omega_{n}) \, \Sigma(\mathbf{k},\omega_{n}) \,  G_{0}(\mathbf{k},\omega_{n}) \,\, ,         \label{BEC-G-1}
\end{equation}

\noindent
and consistently approximate the self-energy (\ref{pairing-self-energy}) in the form:
\begin{equation}
\Sigma (\mathbf{k},\omega_{n}) \simeq - G_{0}(-\mathbf{k},-\omega_{n}) \, k_{B} T \sum_{\nu} e^{i \Omega_{\nu \eta}} 
\int \! \frac{d\mathbf{q}}{(2\pi)^{3}} \, \Gamma^{0}(\mathbf{q},\Omega_{\nu}) \,\, .                        \label{BEC-pairing-self-energy}
\end{equation}

\noindent
In this way, we obtain for the density (\ref{fermion-density}):
\begin{equation}
n \simeq 2 \int \, \frac{d\mathbf{k}}{(2\pi)^{3}} \frac{1}{e^{\xi(\mathbf{k})/(k_{B}T)} + 1} - 
2  \int \! \frac{d\mathbf{q}}{(2 \pi)^{3}} \, k_{B} T \sum_{\nu}  
\frac{e^{i\Omega_{\nu}\eta}}{i\Omega_{\nu} - \mathbf{q}^{2}/(4m) + \mu_{B}}                              \label{BEC-fermion-density}
\end{equation}

\noindent
where use has been made of the result
\begin{equation}
\int \! \frac{d\mathbf{k}}{(2 \pi)^{3}} \, k_{B} T \sum_{n} \,G_{0}(\mathbf{k},\omega_{n})^{2} \, G_{0}(\mathbf{-k},-\omega_{n}) \,
\simeq \, - \, \frac{m^{2}a_{F}}{8\pi}                                                                                                \label{triple-G-o}
\end{equation}

\noindent
which is valid when $\mu \simeq \ - \varepsilon_{0}/2$ is the largest energy scale in the problem.
Under these circumstances, the first term on the right-hand side of Eq.(\ref{BEC-fermion-density}) is strongly suppressed
by the smallness of the fugacity $e^{\mu/(k_{B}T)}$, while the second term therein represents the density $n_{B}$ of a 
non-interacting system of (composite) bosons with chemical potential $\mu_{B}$, yielding eventually $n \simeq 2 n_{B}$.
With the inclusion of pairing fluctuations, the density equation (\ref{fermion-density}) thus reproduces the standard result for
the Bose-Einstein condensation temperature $k_{B} T_{\mathrm{BEC}} = 3.31 n_{B}^{2/3}/m_{B}$ where $m_{B} = 2m$ is the mass of a composite boson.

Note further that, if only the first term on the right-hand side of Eq.(\ref{BEC-fermion-density}) were retained, one would get
for the chemical potential:
\begin{equation}
\frac{\mu}{k_{B} T} \, \simeq \, \ln\left[\frac{n}{2} \left(\frac{2\pi}{m k_{B} T}\right)^{3/2}\right]  \,\,      \label{classical-mu}
\end{equation}

\noindent
which coincides with the classical expression at temperature $T$ \cite{FW}. 
Setting in this expression $\mu \simeq - \varepsilon_{0}/2$ and $T=T_{c}$, the value (\ref{T-c-strong-coupling}) for the critical temperature $T_{c}$ is readily recovered.

Quite generally at any coupling across the BCS-BEC crossover, the critical temperature is obtained from the normal phase by enforcing in Eq.(\ref{pairing-propagator}) the instability condition $1/\Gamma^{0}(\mathbf{q}=0,\Omega_{\nu}=0)=0$, in conjunction with the density equation (\ref{fermion-density}).
The resulting values for $T_{c}$ are plotted for the homogeneous (h) and trapped (t) cases in Fig.\ref{figure3}, where they correspond
the lower dashed and full lines, respectively.
In both cases, in the temperature window between $T_{c}$ and $T^{*}$ composite bosons are formed but not yet condensed.

In the context of the BCS-BEC crossover, pairing fluctuations in the normal phase were first considered by Nozi\`{e}res and Schmitt-Rink (NSR) \cite{NSR}, with the purpose of obtaining a sensible extrapolation of the critical temperature from the BCS to the BEC limit (in
Ref.\cite{NSR} the density equation was obtained by an alternative procedure via the thermodynamic potential).
It was later remarked in Ref.\cite{Serene-89} that the NSR procedure corresponds to a t-matrix theory in which one keeps only the 
lowest-order terms of Eq.(\ref{BEC-G-1}) for all couplings and not just in the BEC limit.
In practice, differences between the numerical results, obtained alternatively by the NSR procedure or by the approach based on Eqs.(\ref{pairing-self-energy})-(\ref{fermion-density}) where the expansion (\ref{BEC-G-1}) is avoided, remain sufficiently small even in the unitary region.

The approach for the normal phase based on Eqs.(\ref{pairing-self-energy})-(\ref{fermion-density}) was considered in Ref.\cite{PPSC-02} to study fermionic single-particle properties above $T_{c}$ in the homogeneous case, and later extended 
to consider the effects of a trap.
Owing to the presence of two bare fermion propagator $G_{0}$ in the particle-particle bubble of Eq.(\ref{pairing-propagator}), this
approach is sometimes referred to as the ``$G_{0}$-$G_{0}$ t-matrix''.
This is to distinguish it from alternative t-matrix approaches, notably: 
(i) The ``$G$-$G_{0}$ t-matrix'' approach \cite{Levin-review} where one bare $G_{0}$ and one self-consistent $G$ enter the particle-particle bubble defining the pair propagator, while a bare $G_{0}$ is kept in the definition of the fermionic self-energy 
(cf. Eq.(\ref{pairing-self-energy}));
(ii) The ``$G$-$G$ t-matrix'' approach \cite{Zwerger-07} where all single-particle Green's functions are self-consistent ones.
These alternative approaches were both utilized recently to study the fermionic single-particle spectral function in the normal phase \cite{Levin-09,Zwerger-09}.
[It should be mentioned in this context that a t-matrix approach formally similar to the $G_{0}$-$G_{0}$ one was proposed in 
Ref.\cite{Sachdev-08}, where the bare value of the chemical potential for the non-interacting Fermi gas was inserted in 
the self-energy in the spirit of a $1/N$ expansion.]

\begin{figure}[htbp]
\begin{center}
\includegraphics[width=10.5cm,angle=0]{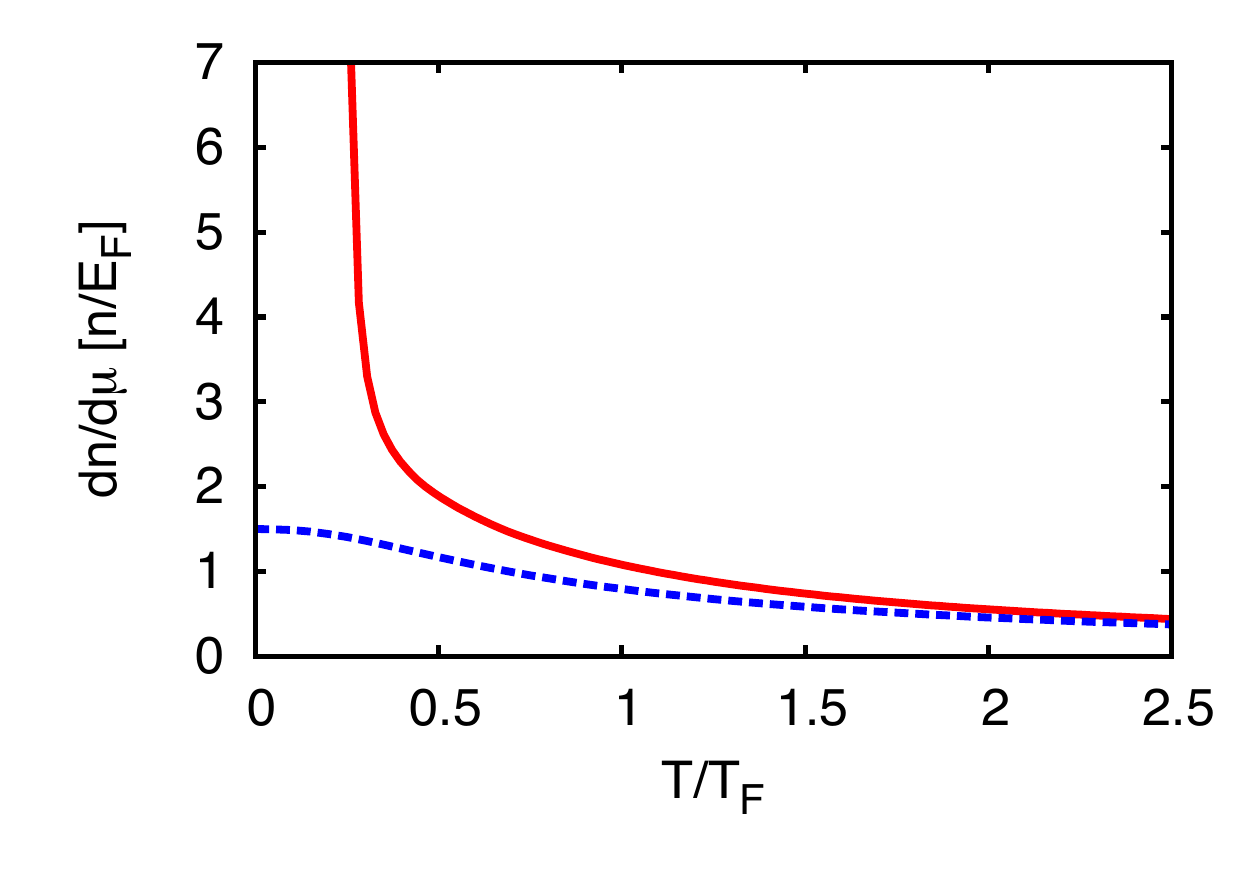}
\caption{The isothermal compressibility $d n / d \mu$ (in units of $n/E_{F}$) vs $T/T_{F}$, as obtained at unitarity from 
              the pairing-fluctuation approach based on Eqs.(\ref{pairing-self-energy})-(\ref{fermion-density}), is shown  
              to diverge at $T_{c}$ (full line). 
              In contrast, the curve corresponding to the free Fermi gas goes smoothly through $T_{c}$ (dashed line).}
\label{figure5}
\end{center}
\end{figure}

While the $G$-$G_{0}$ and $G$-$G$ t-matrix approaches have been implemented according to their strict definitions, the
pairing-fluctuation approach of Eqs.(\ref{pairing-self-energy})-(\ref{fermion-density}) can be allowed to retain the original flexibility of the diagrammatic fermionic structure which is ``modular'' in nature.
In this sense, it can be progressively improved by including additional self-energy corrections which are regarded important, especially 
in the BCS and BEC regimes where the approximations can be controlled. 
This implies, in particular, that the pair propagator in the expression (\ref{pairing-self-energy}) can be dressed via ``bosonic'' self-energy
insertions, which lead, for instance, to the Gorkov and Melik-Barkudarov corrections \cite{GMB-61} on the BCS side
and to the Popov theory for composite bosons \cite{PS-05} on the BEC side.
Consideration of the latter is expected to be especially important on physical grounds, since it effectively introduces a repulsive interaction among the composite bosons which ensures, in particular, the stability of the system under compression.

One major shortcoming of the pairing-fluctuation approach of Eqs.(\ref{pairing-self-energy})-(\ref{fermion-density}) is, in fact, that it leads 
to a diverging compressibility when the temperature is lowered down to $T_{c}$ from the normal phase.
This behavior is shown in Fig.\ref{figure5} at unitarity, and can be ascribed to the fact that the pair propagator (\ref{pairing-propagator}) corresponds to \emph{non-interacting} composite bosons.

The price one has to pay, for setting up theoretical improvements over and above the pairing-fluctuation approach discussed in 
the present Section, is the unavoidable increase of their numerical complexity when calculating physical quantities.
Some of these improvements will be discussed in the next Section.

The above pairing-fluctuation approach can, in addition, be extended to the superfluid phase below $T_{c}$,
whereby the pair propagator acquires a matrix structure that maps onto the bosonic normal and anomalous propagators 
within the Bogoliubov theory \cite{APS-03,PPS-04}. 
This extension (together with its Popov refinement \cite{PS-05}) will also be considered in the next Section.

 
\section{Bogoliubov and Popov approaches, and the residual boson-boson interaction}
\label{sec:Bogoliubov-Popov}

A pairing-fluctuation approach was implemented on physical grounds below $T_{c}$ in Refs.\cite{APS-03,PPS-04}, by adopting 
a fermionic self-energy in the broken-symmetry phase that represents fermions coupled to superconducting fluctuations in weak coupling and to bosons described by the Bogoliubov theory in strong coupling.
This approach has allowed for a systematic study of the BCS-BEC crossover in the temperature range $0<T<T_{c}$. 

A diagrammatic theory for the BCS-BEC crossover below $T_{c}$ was actually first proposed by Haussmann \cite{Haussmann-93}, by extending the self-consistent t-matrix approximation to the broken-symmetry phase.
While the ensuing coupled equations for the chemical potential and order parameter were initially solved at $T_{c}$ only,
an improved version of this self-consistent theory was recently implemented for the whole thermodynamics of the BCS-BEC crossover 
\cite{Zwerger-07}.
We postpone an explicit comparison with this alternative approach to Section~\ref{sec:thermo-dyna}, where a selection of numerical results will be presented.

By the approach of Refs.\cite{APS-03,PPS-04}, the pair propagator in the broken-symmetry phase has the following matrix structure:
\begin{equation}
\left( 
\begin{array}{cc}
\Gamma_{11}(q)&\Gamma_{12}(q)\\
\Gamma_{21}(q)&\Gamma_{22}(q)\end{array}\right)
\, = \, 
\frac{\left( 
\begin{array}{cc} 
A(-q)&B(q) \\ 
B(q)&A(q) \end{array} \right)} 
{A(q)  A(-q) - B(q)^{2}}                                                                                            \label{Gamma-solution}
\end{equation}  

\noindent
where 
\begin{eqnarray}
- A(q) & = & \frac{m}{4\pi a_F} +  \int \! \frac{d {\mathbf k}}{(2\pi)^{3}} 
\left[ k_{B} T \, \sum_{n} {\mathcal G}_{11}(k+q) {\mathcal G}_{11}(-k) \, - \, \frac{m}{|{\bf k}|^2} \right]     \label{A-definition} \\
B(q) & = & \int \! \frac{d {\mathbf k}}{(2\pi)^{3}} \, k_{B} T \, \sum_{n} \,
{\mathcal G}_{12}(k+q) \,{\mathcal G}_{21}(-k)           \,\, .                                                                         \label{B-definition}
\end{eqnarray}

\begin{figure}[htbp]
\begin{center}
\includegraphics[width=8.0cm,angle=0]{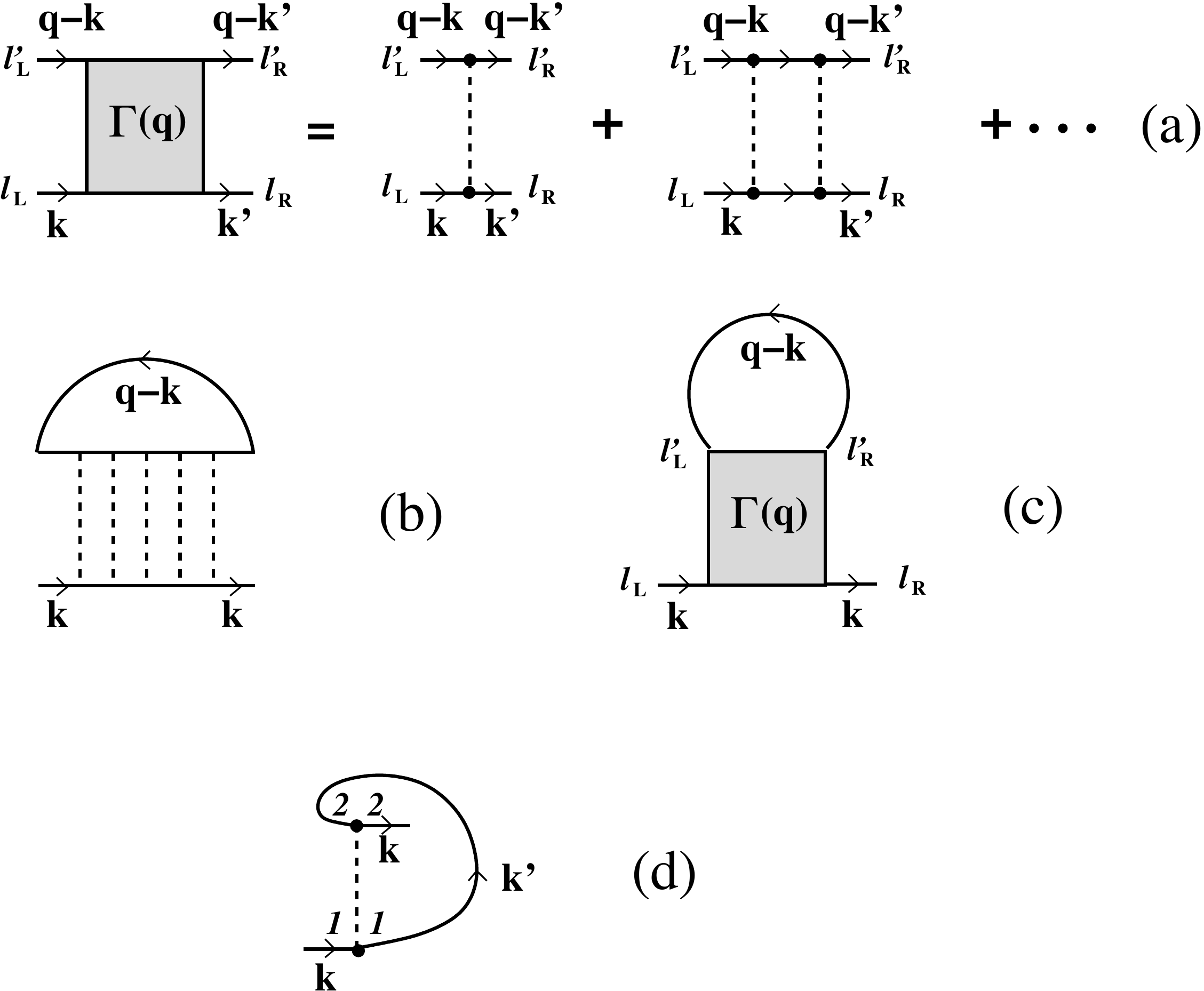}
\caption{Single-particle fermionic self-energy for the broken-symmetry phase (panel (c)), expressed in terms of the pair 
              propagator $\Gamma$ with Nambu structure (panel (a)).
              The BCS contribution to the self-energy is shown in panel (d), and the corresponding self-energy for the normal-phase 
              of Fig.\ref{figure4} is also reported in panel (b) for comparison. (Reproduced from Fig.1 of Ref.\cite{PPS-04}.)}
\label{figure6}
\end{center}
\end{figure}

\noindent
This structure is represented diagrammatically in Fig.\ref{figure6}a, where only combinations with $\ell_{L} = \ell'_{L}$ and
$\ell_{R} = \ell'_{R}$ survive the regularization we have adopted for the potential [cf. Eq.(\ref{aF-vs-v0})].
It represents an approximation to the Bethe-Salpeter equation for the fermionic two-particle GreenÕs function in the particle-particle channel.
In the above expressions, $q=({\mathbf q},\Omega_{\nu})$ and $k=({\mathbf k},\omega_{n})$ are four-vectors, and 
\begin{eqnarray}
{\mathcal G}_{11}({\mathbf k},\omega_n) \, & = & \, - \frac{\xi({\mathbf k}) + i \omega_n}
{E({\mathbf k})^2 + \omega_n^2} \, = \, - \, {\mathcal G}_{22}(-{\mathbf k},-\omega_n)                            \nonumber  \\
{\mathcal G}_{12}({\mathbf k},\omega_n) \, & = & \, \frac{\Delta}
{E({\mathbf k})^2 + \omega_n^2} \, = \, {\mathcal G}_{21}({\mathbf k},\omega_n)                                    \label{BCS-Green-function}
\end{eqnarray}

\noindent
are the BCS single-particle Green's functions in Nambu notation \cite{Schrieffer}, with $E({\mathbf k})=\sqrt{\xi({\mathbf k})^{2}+\Delta^{2}}$ for an isotropic ($s$-wave) order parameter $\Delta$ (which we take to be real without loss of generality).  

In analogy to what was done for obtaining the expression (\ref{bosonic-propagator}) in the strong-coupling limit, one can
show that the pair propagator (\ref{Gamma-solution}) reduces in the same limit to the following expressions:
\begin{equation}
\Gamma_{11}(q) \, = \, \Gamma_{22}(-q) \, \simeq \, \frac{8 \pi}{m^{2} a_{F}} \,
\frac{\mu_{B} + i \Omega_{\nu} + {\mathbf q}^{2}/(4m)}{E_{B}({\mathbf q})^{2} - (i \Omega_{\nu})^{2}}     
                                                                                                                                                       \label{Gamma-11-approx}
\end{equation}

\noindent
and
\begin{equation}
\Gamma_{12}(q) \, = \, \Gamma_{21}(q) \, \simeq \, \frac{8 \pi}{m^{2} a_{F}} \,
\frac{\mu_{B}}{E_{B}({\mathbf q})^{2} - (i \Omega_{\nu})^{2}}  \,\, ,                                                \label{Gamma-12-approx}
\end{equation}

\noindent
where
\begin{equation}
E_{B}({\mathbf q}) \, = \, \sqrt{ \left( \frac{{\mathbf q}^{2}}{2 m_{B}} + \mu_{B}\right)^{2} - \mu_{B}^{2}}
                                                                                                                                                    \label{Bogoliubov-disp}
\end{equation}

\noindent
has the form of the Bogoliubov dispersion relation \cite{FW}, $\mu_{B} = \Delta^{2}/(4 |\mu|) = 2 \mu + \varepsilon_{0}$ being 
the corresponding value of the bosonic chemical potential.
Apart from the overall factor $- 8 \pi/(m^{2} a_{F})$ (and a sign difference in the off-diagonal component \cite{APS-03}), 
the expressions (\ref{Gamma-11-approx}) and (\ref{Gamma-12-approx}) coincide, respectively, with the normal and anomalous
non-condensate bosonic propagators within the Bogoliubov approximation \cite{FW}.

For any coupling, in Ref.\cite{PPS-04} the corresponding fermionic self-energy was taken of the form:
\begin{eqnarray}
\Sigma_{11}(k) & = & - \Sigma_{22}(-k) =  
- k_{B} T \sum_{\nu} \int \! \frac{d\mathbf{q}}{(2\pi)^{3}} \, \Gamma_{11}(\mathbf{q},\Omega_{\nu}) \, 
{\mathcal G}_{11}(\mathbf{q}-\mathbf{k},\Omega_{\nu}-\omega_{n})                                                      \nonumber\\
\Sigma_{12}(k) & = &  \Sigma_{21}(k)  =  - \Delta                                                                                    \label{Sigma-broken}
\end{eqnarray}

\noindent
where $\Sigma_{11}$ is shown diagrammatically in Fig.\ref{figure6}c  (with $\ell_{L} = \ell'_{L} = \ell_{R} = \ell'_{R} =1$) and 
$\Sigma_{12}$ in Fig.\ref{figure6}d.

With this choice of the self-energy, the fermionic propagator is then obtained by solving Dyson's equation in matrix form:
\begin{equation}
\left( \begin{array}{cc} G_{11}^{-1}(k) & G_{12}^{-1}(k)  \\ G_{21}^{-1}(k) & G_{22}^{-1}(k) \end{array} \right)
= \left( \begin{array}{cc} G_{0}(k)^{-1} & 0 \\ 0 & - G_{0}(-k)^{-1} \end{array} \right) -               
\left( \begin{array}{cc} \Sigma_{11}(k) & \Sigma_{12}(k) \\ \Sigma_{21}(k) & \Sigma_{22}(k) \end{array} \right) \, .           
                                                                                                                                                             \label{Dyson-equation}         
\end{equation}

\noindent
Note that the BCS expressions (\ref{BCS-Green-function}) for the single-particle GreenÕs functions result by neglecting 
in Eqs.(\ref{Dyson-equation}) the diagonal self-energy terms associated with pairing fluctuations.

By the approach of Ref.\cite{PPS-04}, the ``normal'' propagator $G_{11}$ is inserted in the density equation
\begin{equation}
n \, = \, 2 \, k_{B} T \sum_{n} e^{i \omega_{n} \eta} \int \, \frac{d\mathbf{k}}{(2\pi)^{3}} \, G_{11}(\mathbf{k},\omega_{n})
                                                                                                                                                                  \label{broken-fermion-density}
\end{equation} 

\noindent
which replaces Eq.(\ref{fermion-density}) below $T_{c}$, while in the gap equation 
\begin{equation}
\Delta \, = \, - \, v_{0} \, k_{B} T \sum_{n} \int \, \frac{d\mathbf{k}}{(2\pi)^{3}} \, \, {\mathcal G}_{12}({\mathbf k},\omega_{n})  
                                                                                                                                                                      \label{Delta-G-12}
\end{equation}

\noindent
 the BCS ``anomalous'' propagator (\ref{BCS-Green-function}) is maintained (albeit with modified numerical values of the chemical potential and order parameter that result from the simultaneous solution of Eqs.(\ref{broken-fermion-density}) and (\ref{Delta-G-12})).
This ensures that the bosonic propagators (\ref{Gamma-solution}) remain \emph{gapless}.
 
It is instructive to consider once more the BEC limit, whereby the diagonal part of the self-energy acquires the following approximate form \cite{PPS-04}:
\begin{equation}
\Sigma_{11}({\mathbf k},\omega_{n}) \, \simeq \, \frac{8 \pi}{m^{2} a_{F}} \,\, \frac{1}{i\omega_{n} + \xi({\mathbf k})} \,\, n'_{B}(T) \,\, .
                                                                                                                                                             \label{Sigma-11-n-prime}
\end{equation}

\noindent
Here,
\begin{equation}
n'_{B}(T) = \int \! \frac{d {\mathbf q}}{(2 \pi)^{3}} 
\left[ u_{B}^{2}({\mathbf q}) b(E_{B}({\mathbf q})) - v_{B}^{2}({\mathbf q}) b(-E_{B}({\mathbf q})) \right]
                                                                                                                                                             \label{noncondensate-density}
\end{equation}

\noindent
represents the bosonic \emph{noncondensate density}, with the Bose distribution $b(x) = \{ \exp [x/(k_{B}T)] - 1 \}^{-1}$ and the standard bosonic factors of the Bogoliubov transformation \cite{FW}:
\begin{equation}
v_{B}^{2}({\mathbf q}) \, = \, u_{B}^{2}({\mathbf q}) - 1 \, = \, \frac{\frac{{\mathbf q}^{2}}{2m_{B}} \, + \mu_{B} \, - \,
E_{B}({\mathbf q})}{2 E_{B}({\mathbf q})}  \,\, .                                                                                      \label{u-v-Bogoliubov}
\end{equation}

\noindent
In this case, solution of the Dyson's equation (\ref{Dyson-equation}) yields:
\begin{equation}
G_{11}({\mathbf k},\omega_{n}) \, \simeq \, \frac{1}{i\omega_{n} \, - \, \xi({\mathbf k}) \, - \,
\frac{\Delta^{2} \, + \, \Delta_{\mathrm{pg}}^{2}}{i\omega_{n} \, + \, \xi({\mathbf k})}}                                         \label{G-11-strong-coupling}
\end{equation}

\noindent
with the notation $\Delta_{\mathrm{pg}}^{2} = 8 \pi n'_{B}(T)/(m^{2} a_{F})$.
When inserted into the density equation (\ref{broken-fermion-density}) the above expression gives:
\begin{equation}
n \, \simeq \, \frac{m^{2} \, a_{F}}{4 \pi} \, \left( \Delta^{2} \, + \, \Delta_{\mathrm{pg}}^{2} \right) \, = \, 2 \left(  n_{0}(T) \, + \, n'_{B}(T) \right)    
                                                                                                                                                                   \label{BEC-final-n}
\end{equation}

\noindent
where the \emph{condensate density} $n_{0}(T)$ is identified via $\Delta^{2} = 8 \pi n_{0}(T)/(m^{2} a_{F})$.

It is relevant to comment at this point on the value of the scattering length $a_{B}$, which results in the BEC limit of the above approach 
from the residual interaction between composite bosons.
This value is obtained, for instance, by manipulating the gap equation (\ref{Delta-G-12}) in this limit, yielding:
\begin{equation}
\frac{\Delta^{2}}{4|\mu|} \simeq 2 \left( \sqrt{2 |\mu| \varepsilon_{0}} \, - \, 2 |\mu| \right) \simeq \mu_{B} \,\, . 
                                                                                                                                                           \label{BCS-gap-equation-sc}
\end{equation}

\noindent
With the relation between $\Delta^{2}$ and $n_{0}$ utilized in Eq.(\ref{BEC-final-n}) and the asymptotic result 
$|\mu| \simeq (2 m a_{F}^{2})^{-1}$,
the expression (\ref{BCS-gap-equation-sc}) can be cast in the form $\mu_{B} = 4 \pi a_{B} n_{0}/m_{B}$ that corresponds to the value of the Bogoliubov theory with $a_{B} = 2 a_{F}$.

\begin{figure}[htbp]
\begin{center}
\includegraphics[width=7.0cm,angle=0]{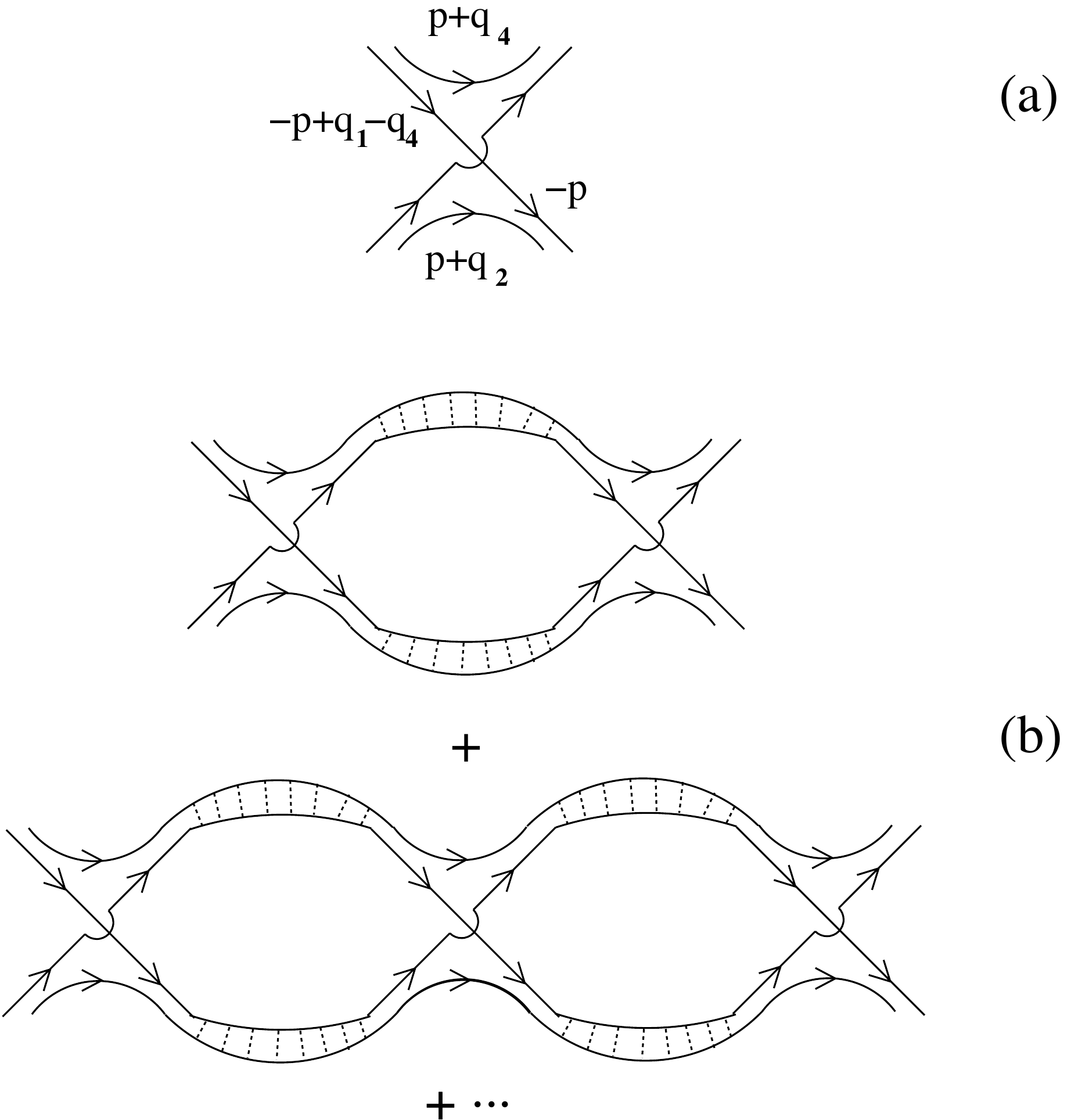}
\caption{(a) Effective boson-boson interaction $\bar{u}_{2}$.
              (b) Additional terms associated with the t-matrix $\bar{t}_{B}$ for composite bosons.
               Light lines stand for free-fermion propagator and broken lines for fermionic interaction potential. 
               Spin labels are not shown explicitly. (Reproduced from Fig.2 of Ref.\cite{PS-05}.)}
\label{figure7}
\end{center}
\end{figure}

This result can also be interpreted diagrammatically as being associated with the lowest-order (Born approximation) value for the effective
boson-boson interaction \cite{PS-00}.
This is represented in Fig.\ref{figure7}a and can be obtained from the following expression where all bosonic four-momenta $q_{i}$ ($i=1,\cdots,4$) vanish:
\begin{eqnarray}
\bar{u}_{2}(0,0,0,0) & = & k_{B} T \, \sum_{n} \, \int \! \frac{d \mathbf{p}}{(2\pi)^{3}} \, G_{0}(p)^{2} \,\,  G_{0}(-p)^{2}  \nonumber \\
& \simeq & \int \!\! \frac{d \mathbf{p}}{(2\pi)^{3}} \, \frac{1}{4 \xi(\mathbf{p})^{3}}
   \simeq \left( \frac{m^{2} a_{F}}{8 \pi} \right)^{2}  \left( \frac{4 \pi a_{F}}{m} \right)                                                       \label{u-0-sc}
\end{eqnarray}

\noindent
with the last line holding in the BEC limit.
Apart from the overall factor $(m^{2} a_{F}/(8 \pi))^{2}$ (that compensates for the presence of the factor $- 8 \pi/(m^{2} a_{F})$ in
the expression (\ref{bosonic-propagator}) of the free-boson propagator), the result (\ref{u-0-sc}) is indeed consistent with a residual 
bosonic interaction corresponding to $a_{B} = 2 a_{F}$. 

The correct value for $a_{B}$ ($ = 0.6 a_{F}$), which includes all possible scattering processes between two composite bosons in isolation, was originally determined in Ref.\cite{PSS-2004} from the exact solution of the Schr\"{o}dinger equation for dimer-dimer elastic scattering, and later confirmed in Ref.\cite{KC-2005} through a completely diagrammatic treatment at zero density.
In this context, even before this exact result was available, it was shown in Ref.\cite{PS-00} that the scattering processes corresponding to the t-matrix diagrams for composite bosons (the lowest ones of which are depicted in Fig.\ref{figure7}b) lead by themselves to a considerable reduction of the value of $a_{B}$ ($ \simeq 0.75 a_{F}$) starting from the value $a_{B} = 2 a_{F}$ of the Born approximation.
As a matter of fact, the complete diagrammatic treatment of Ref.\cite{KC-2005} (that yields the exact value $a_{B} = 0.6 a_{F}$) adds to 
the diagram of Fig.\ref{figure7}a all other additional (zero density) processes which are irreducible with respect to the propagation of two composite bosons, and then uses the result in the place of the diagram of Fig.\ref{figure7}a as the new kernel of the integral equation 
depicted in Fig.\ref{figure7}b.

The above considerations suggest us a way to improve on the Bogoliubov approximation for composite bosons, in order to
include the diagrammatic contributions leading to a refined value of $a_{B}$ with respect to the Born approximation
(in the following, we shall limit ourselves to recovering the value $a_{B} \simeq 0.75 a_{F}$ in the BEC limit).
To this end, we first approximately obtain the pair propagators $\mathbf{\Gamma}_{B}$ for any value of the fermionic coupling, by adopting the following Dyson's type equation in matrix form \cite{PS-05} in the place of the expressions (\ref{Gamma-solution}): 
\begin{equation}
\mathbf{\Gamma}_{B}(q) \, = \, \mathbf{\Gamma}_{B}^{0}(q) \, + \, 
\mathbf{\Gamma}_{B}^{0}(q) \, \mathbf{\Sigma}_{B}(q) \, \mathbf{\Gamma}_{B}(q) \,\, .                   \label{boson-Dyson-equation}
\end{equation}

\noindent
Here, $\mathbf{\Gamma}_{B}^{0}(q)$ is the free-boson propagator with inverse
\begin{equation}
\mathbf{\Gamma}_{B}^{0}(q)^{-1} = \left( \begin{array}{cc} 
\Gamma^{0}(q)^{-1} &   0   \\
0 & \Gamma^{0}(-q)^{-1}
\end{array} \right)                                                                                                                        \label{inverse-free-boson}
\end{equation}

\noindent
where $\Gamma^{0}(q)^{-1}$ is given by Eq.(\ref{pairing-propagator}), and
\begin{equation}
\mathbf{\Sigma}_{B}(q) \, = \, \Delta^{2} \, \left( \begin{array}{cc} 
- 2 \, \bar{u}_{2}(0,q,0,q) &  \bar{u}_{2}(0,0,-q,q)   \\
\bar{u}_{2}(0,0,-q,q) &  - 2 \, \bar{u}_{2}(0,q,0,q)
\end{array} \right)                                                                                                                            \label{Bogoliubov-self-energy-cb}
\end{equation}

\noindent
is the bosonic self-energy within the Bogoliubov approximation, which contains two degenerate forms of the effective boson-boson
interaction [cf. Fig.\ref{figure7}a]:
\begin{equation}
\bar{u}_{2}(q_{1},q_{2},q_{3},q_{4}) = k_{B} T \sum_{n}  \int \!\!  \frac{d\mathbf{p}}{(2\pi)^{3}}  
G_{0}(-p) G_{0}(p+q_{2}) G_{0}(-p+q_{1}-q_{4}) G_{0}(p+q_{4}) .                                             \label{u-q} 
\end{equation}

\noindent
To guarantee the ladder propagators $\mathbf{\Gamma}_{B}(q)$ of Eq.(\ref{boson-Dyson-equation}) to be gapless when $q=0$ for any 
value of the fermionic coupling, we impose the condition:
\begin{equation}
\Gamma^{0}(q=0)^{-1} - \Sigma_{B}^{11}(q=0) - \Sigma_{B}^{12}(q=0) =  0                                \label{HP-Bogoliubov-cb}
\end{equation}
which plays the role in the present context of the Hugenholtz-Pines theorem for point-like bosons \cite{FW}.
The Bogoliubov approximation for the composite bosons in the BEC limit with an improved value of $a_{B}$ then results \cite{PS-00}, by replacing in Eq.(\ref{Bogoliubov-self-energy-cb}) the boson-boson interaction $\bar{u}_{2}$ with the following expression of the t-matrix for composite bosons (cf. Fig.\ref{figure7}b):
\begin{eqnarray}
&& \bar{t}_{B}(q_{1},q_{2},q_{3},q_{4}) =  \bar{u}_{2}(q_{1},q_{2},q_{3},q_{4})
- k_{B} T \, \sum_{\nu_{5}} \int \!\! \frac{d \mathbf{q}_{5}}{(2\pi)^{3}}                                            \label{bosonic-t-matrix}  \\
&& \times \bar{u}_{2}(q_{1},q_{2},q_{5},q_{1}+q_{2}-q_{5}) \Gamma^{0}(q_{5}) \Gamma^{0}(q_{1}+q_{2}-q_{5}) \,
 \bar{t}_{B}(q_{1}+q_{2}-q_{5},q_{5},q_{3},q_{4})  .                                                                      \nonumber
\end{eqnarray}

\begin{figure}[htbp]
\begin{center}
\includegraphics[width=5.0cm,angle=0]{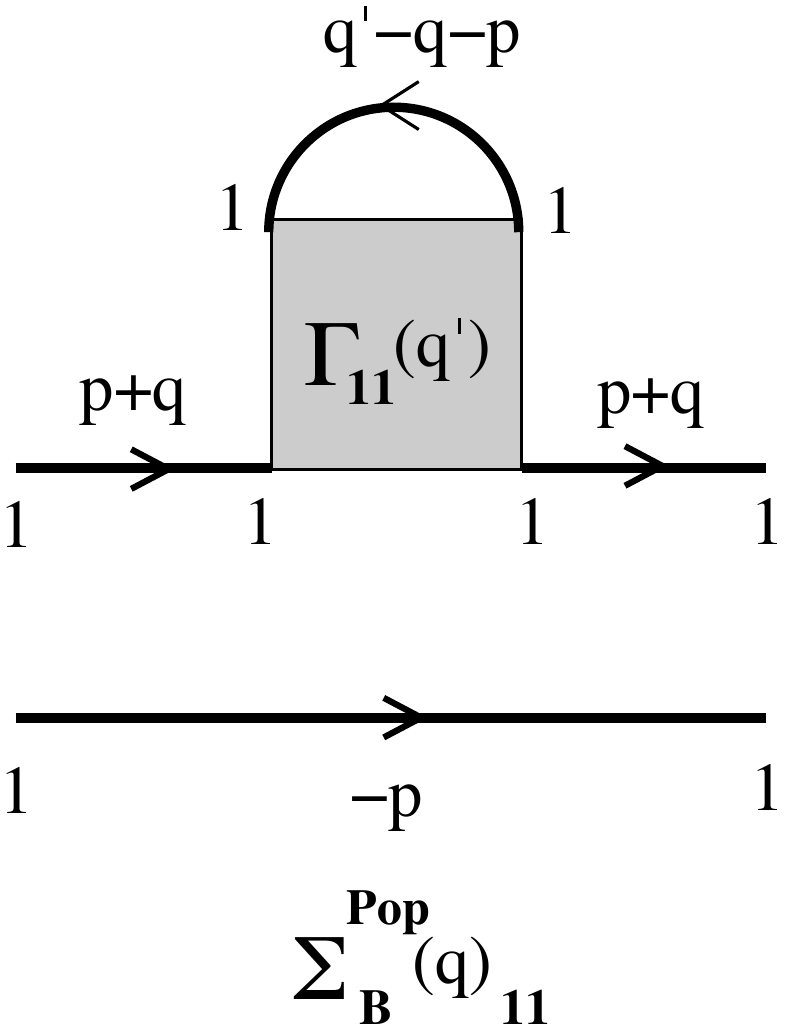}
\caption{Graphical representation of the Popov self-energy for composite bosons, that results upon dressing the upper fermionic line in the 
              particle-particle channel. An analogous dressing done for the lower fermionic line accounts for the factor of two
              in Eq.(\ref{Sigma-Popov-comp-bosons-full}). (Reproduced from Fig.4 of Ref.\cite{PS-05}.)}
\label{figure8}
\end{center}
\end{figure} 

Further improvements can be implemented by replacing the pair propagators (\ref{Gamma-solution}) with more refined descriptions of composite bosons in the BEC limit, and then using these improved descriptions throughout the BCS-BEC crossover to modify the fermionic single-particle self-energy accordingly.
An example is the so-called Popov approximation for composite bosons \cite{PS-05}, whereby the bosonic self-energy 
represented diagrammatically in Fig.\ref{figure8} is employed to modify the original Bogoliubov propagators (\ref{Gamma-solution}).
In the broken-symmetry phase, this bosonic self-energy has the form: 
\begin{eqnarray}
\Sigma_{B}^{\rm{Pop}}(q)_{11} = - 2 \, k_{B} T \, \sum_{n} \int \! \frac{d \mathbf{p}}{(2\pi)^{3}} \,\,
k_{B} T \, \sum_{\nu'} \int \!\! \frac{d \mathbf{q'}}{(2\pi)^{3}}                                                                       \nonumber \\
\times \, G_{11}(p+q)^{2} \, G_{11}(-p) \, G_{11}(q'-q-p) \, \Gamma_{11}(q') \, .                                \label{Sigma-Popov-comp-bosons-full}
\end{eqnarray}

The Popov propagators for composite bosons are then obtained as follows, in terms of the corresponding Bogoliubov propagators 
(\ref{Gamma-solution}):
\begin{equation}
\left( 
\begin{array}{cc} \Gamma^{\rm{Pop}}_{11}(q) & \Gamma^{\rm{Pop}}_{12}(q) \\ \Gamma^{\rm{Pop}}_{21}(q) & \Gamma^{\rm{Pop}}_{22}(q) \end{array} \right)
=  \frac{\left( 
\begin{array}{cc} 
A(-q)  -  \Sigma_{B}^{\rm{Pop}}(-q)_{11} & B(q) \\ B(q) & A(q)  -  \Sigma_{B}^{\rm{Pop}}(q)_{11}\end{array} \right)}
{[A(q) - \Sigma_{B}^{\rm{Pop}}(q)_{11}]  [A(-q) - \Sigma_{B}^{\rm{Pop}}(-q)_{11}]  -  B(q)^{2}}                           \label{Popov-propagators}
\end{equation}

\noindent
where $A(q)$ and $B(q)$ are given by Eqs.(\ref{A-definition}) and (\ref{B-definition}), in the order.
The propagators (\ref{Popov-propagators}) are gapless provided
\begin{equation}
A(q=0)  - \Sigma_{B}^{\rm{Pop}}(q=0)_{11}  -  B(q=0)  = 0 \, .                                                      \label{Gapless-condition-Popov-cb}
\end{equation}

\noindent
This generalizes to the present context the condition $A(q=0)-B(q=0)=0$ for gapless Bogoliubov propagators, and effectively replaces the gap equation (\ref{Delta-G-12}) for all practical purposes.

In addition, the same treatment that was made above to improve on the relation $a_{B} = 2 a_{F}$ in the BEC limit can be applied here,
by first rewriting the expression (\ref{Sigma-Popov-comp-bosons-full}) in terms of the bare boson-boson interaction (\ref{u-q})
\begin{equation}
\Sigma_{B}^{\rm{Pop}}(q)_{11} \, \simeq \, - \, 2 \, k_{B} T \, \sum_{\nu'} \, \int \! \frac{d \mathbf{q'}}{(2\pi)^{3}} \, 
\bar{u}_{2}(q',q,q',q) \, \Gamma_{11}(q') \, ,                                                                             \label{Sigma-Popov-comp-bosons-full-u-2}
\end{equation}

\noindent
and then replacing $\bar{u}_{2}$ by the t-matrix $\bar{t}_{B}$ for composite bosons of Eq.(\ref{bosonic-t-matrix}).

The final form of the fermionic self-energy is eventually obtained by reconsidering the expressions (\ref{Sigma-broken}), where now the
Popov propagator $\Gamma^{\rm{Pop}}_{11}$ of Eq.(\ref{Popov-propagators}) takes the place of $\Gamma_{11}$ while
$\Delta$ satisfies the condition (\ref{Gapless-condition-Popov-cb}) in the place of the original gap equation.

From a physical point of view, the relevance of the Popov approximation results because it introduces \emph{an effective repulsion} among the composite bosons through the presence of their noncondensate density.
The importance of this repulsion should be especially evident in the normal phase, when the Bogoliubov propagators 
(\ref{Gamma-solution}) reduce to free-boson propagator (\ref{pairing-propagator}) and miss accordingly this residual bosonic interaction.
While commenting on Fig.\ref{figure5} we have already pointed out that this is the reason for a diverging compressibility at $T_{c}$
when only ``bare'' pairing fluctuations are considered.

In the next Section we shall discuss a number of thermodynamic as well as dynamical results obtained by implementing the Popov approximation \emph{in the normal phase} throughout the BCS-BEC crossover (the unitary limit will specifically be considered).
In this case, the (inverse of the) Popov propagator for composite bosons is obtained from the relation 
$\Gamma^{\rm{Pop}}(q)^{-1} = \Gamma^{0}(q)^{-1} - \Sigma_{B}^{\rm{Pop}}(q)$, where $\Sigma_{B}^{\rm{Pop}}$ is given by 
Eq.(\ref{Sigma-Popov-comp-bosons-full-u-2}) with $\Gamma^{0}$ replacing $\Gamma_{11}$.
In addition, to improve on the description of the boson-boson scattering, we shall replace the bare $\bar{u}_{2}$ in
Eq.(\ref{Sigma-Popov-comp-bosons-full-u-2}) by the t-matrix $\bar{t}_{B}$ for composite bosons given by Eq.(\ref{bosonic-t-matrix}). 
In this way, the 2-boson scattering will be dealt with beyond the Born approximation.

In this context, it will be relevant to compare the results obtained by the above approach in the normal phase for thermodynamic and dynamical quantities \cite{PPS-10}, with those obtained by an alternative approach based on a self-consistent t-matrix approximation, as described in Ref.\cite{Zwerger-07} for the thermodynamics and in Ref.\cite{Zwerger-09} for the dynamics of the BCS-BEC crossover, respectively.
[Results for a homogeneous system will only be presented.]
Interest in this comparison is also justified on physical grounds, by considering the different treatments of the effective boson-boson interaction which result from the two approaches.
As remarked already, the Popov approach with $\bar{t}_{B}$ replacing $\bar{u}_{2}$ concentrates on 2-boson scattering beyond the lowest order (Born) approximation, while the self-consistent t-matrix approach includes a sequence of 3, 4, $\cdots$, n-boson scattering processes, where each process is dealt with at the lowest order.
Although these alternative sets of processes (namely, improved 2-boson vs n-boson scattering) can be clearly identified by a diagrammatic analysis in the (BEC) strong-coupling limit \cite{PS-00}, the question of how the relevance of these processes extends to the unitarity limit remains open and can be addressed only via numerical calculations.
This question will be partially addressed in the next Section.

  
\section{Results for thermodynamic and dynamical quantities}
\label{sec:thermo-dyna}

Physical quantities that can be considered for a quantum many-body system are conveniently organized as single- and two-particle properties, and are correspondingly obtained in terms of single- and two-particles Green's functions.
In addition, these properties may refer to the equilibrium state of the system or to excitations over and above this state.
In the first case they can be conveniently obtained within the Matsubara formalism with discrete imaginary frequencies, while in the second 
case a (sometimes nontrivial) analytic continuation to the real frequency axis is required \cite{FW}.
In the present context of a pairing-fluctuation diagrammatic approach to the BCS-BEC crossover, we shall limit ourselves to considering 
the chemical potential and the total energy per particle as examples of thermodynamic properties, and the single-particle spectral 
function as an example of dynamical properties, for which consideration of pairing fluctuations appears especially relevant.

This relevance is most evident in the \emph{normal phase}, because the occurrence of pairing fluctuations acts to extend 
above $T_{c}$ characteristic effects of pairing (notably, what is referred to as the ``pseudogap physics'' associated with the noncondensate density like in Eq.(\ref{G-11-strong-coupling})), effects which would otherwise be peculiar of the broken-symmetry phase below $T_{c}$ only.

\subsection{Thermodynamic properties}
\label{subsec:thermo}

For a homogeneous Fermi gas in the normal phase, the fermionic chemical potential $\mu$ can be obtained from the density equation 
(\ref{fermion-density}) and the total energy per particle from the following expression \cite{FW}:
\begin{equation}
\frac{E}{N} \, = \, \frac{1}{n} \, k_{B} T \, \sum_{n} \, e^{i \omega_{n} \eta}  \int \! \frac{d\mathbf{k}}{(2 \pi)^{3}} \,
\left( \frac{\mathbf{k}^{2}}{2m} \, + \, \mu \, + \, i \omega_{n} \right) G(\mathbf{k},\omega_{n})                     \label{energy-per-particle}
\end{equation}

\noindent 
where $N$ is the total particle number.
In Eqs.(\ref{fermion-density}) and (\ref{energy-per-particle}), different approximations are embodied in different forms of the fermionic single-particle Green's function $G$. 

In particular, we shall consider approximate forms of $G$ obtained within: 
(i) The t-matrix approach given by Eqs.(\ref{pairing-self-energy})-(\ref{dressed-fermion-propagator}); 
(ii) Its further simplification (sometimes referred to as the Nozi\`{e}res-Schmitt-Rink (NSR) approximation) whereby the fermionic propagator is expanded like in Eq.(\ref{BEC-G-1}) for any coupling (and not just in the BEC limit);
(iii) The Popov approach with an improved description of the boson-boson scattering as discussed in 
Section~\ref{sec:Bogoliubov-Popov};
(iv) The fully self-consistent (sc) t-matrix approach of Ref.\cite{Zwerger-07} that was mentioned in 
Section~\ref{sec:pairing-fluctuations}. 

\begin{figure}[htbp]
\begin{center}
\includegraphics[width=4.7cm,angle=0]{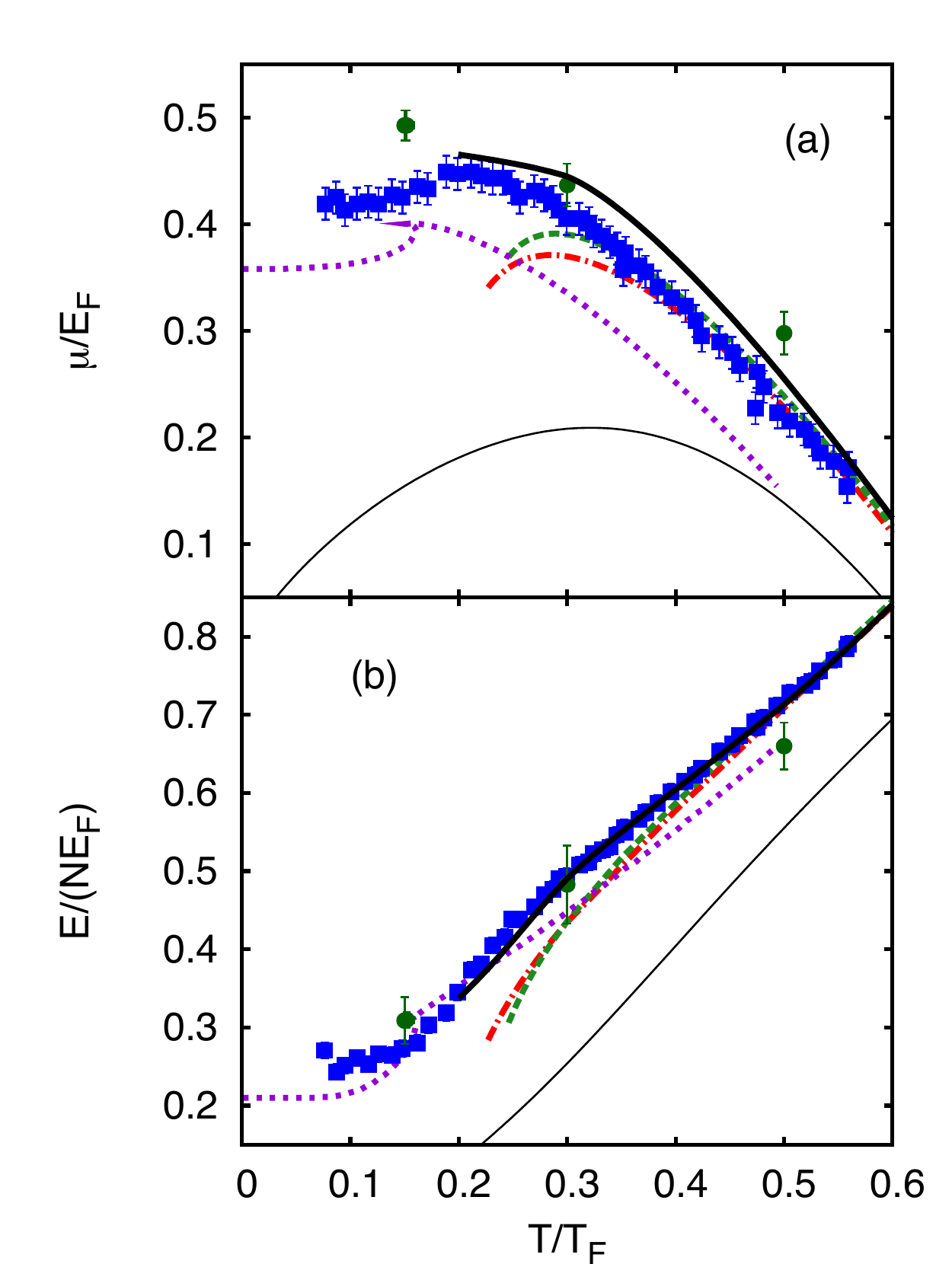}
\caption{(a) Chemical potential and (b) energy per particle (in units of $E_{F}$) vs the temperature (in units of $T_{F}$), as obtained:
               In Ref.\cite{PPS-10} by the t-matrix approach (dashed lines), the NSR approximation (dashed-dotted lines), and the Popov 
               approach (thick full lines); In Ref.\cite{Zwerger-07} by the fully self-consistent t-matrix approach (dotted lines); 
               In Ref.\cite{Bulgac-QMC-2006} (full squares) and in Ref.\cite{Burovski-QMC-2006} (full circles) by QMC calculations.
               Results obtained by the modified virial expansion of Ref.\cite{Salomon-2009} (thin full lines) are also shown for comparison.}
\label{figure9}
\end{center}
\end{figure}

At unitarity, the results of these diagrammatic approaches for $\mu$ and $E/N$ can also be compared with Quantum Monte Carlo (QMC) calculations, which are available over a wide temperature range.
This comparison is shown in Fig.\ref{figure9}. Several features are here apparent. 
At high enough temperatures, all data progressively merge to the t-matrix approach, which is known to become exact in this limit where it reduces to the virial expansion of Beth and Uhlenbeck \cite{Combescot-2006}.
While only minor differences appear between the t-matrix and the NSR approaches (with independent NRS calculations yielding comparable results \cite{Drummond-2006}), the Popov approach is seen to a add positive contribution both to $\mu$ and $E/N$.
This is in line with the expectation that the Popov approach takes into account the residual (repulsive) interaction among composite bosons \cite{PS-05}, which is missed by the t-matrix approach.
In addition, the fully self-consistent t-matrix approach, which at high enough temperature should also asymptotically reduce to the t-matrix approach without self-consistency, shows deviations from the t-matrix and Popov approaches that are more marked in $\mu$ than 
in $E/N$.
No compelling conclusions can, therefore, apparently be drawn by comparing self-consistent vs non-self-consistent pairing-fluctuation approaches as far as the thermodynamic quantities are concerned.
Quite generally, it can be stated that good overall agreement results by comparing QMC calculations with diagrammatic pairing-fluctuation approaches, signifying that the latter are able to capture the relevant physical processes.
Note finally from Fig.\ref{figure9} that the progressively increasing differences between the t-matrix and Popov approaches when lowering the temperature reflects the fact that the divergence of the compressibility resulting from the t-matrix approach (shown in Fig.\ref{figure5}) is suitably cut off by the Popov approach, which yields a finite value for this quantity at $T_{c}$ \cite{PPS-10}.

\begin{table}
\begin{center}
\footnotesize
\begin{tabular}[c]{@{} c@{} | r@{.}l | r@{.}l@{} |r@{.}l@{} |r@{.}l@{} |r@{.}l@{} |r@{.}l@{} |r@{.}l@{} |r@{.}l@{} |r@{.}l@{} |r@{.}l@{}}
& \multicolumn{2}{c|}{t-matrix} & \multicolumn{2}{c|}{NSR} & \multicolumn{2}{c|}{Popov} & \multicolumn{2}{c|}{sc t-matrix} & \multicolumn{2}{c|}{QMC I} & \multicolumn{2}{c|}{QMC II} \\
\hline
$\beta$    & $-0$&$57$~\cite{PPS-10}   & $-0$&$599$~\cite{Drummond-2006} & \multicolumn{2}{c|}{} & $-0$&$64$~ \cite{Zwerger-07}  &    $-0$&$63(5)$~\cite{Bulgac-QMC-2008} & \multicolumn{2}{c|}{}\\
\hline
$\mu(T=0)$ &  $0$&$455$~\cite{PePS-04} &  $0$&$401$~\cite{Drummond-2006} & \multicolumn{2}{c|}{} &  $0$&$36$~ \cite{Zwerger-07}  & $\sim0$&$42$~\cite{Bulgac-QMC-2006} & \multicolumn{2}{c|}{}\\
\hline
$T_c$      &  $0$&$243$~\cite{PPS-10}  &  $0$&$225$~\cite{PPS-10}  & $0$&$199$~\cite{PPS-10}   &  $0$&$160$~ \cite{Zwerger-07} &    $<0$&$15$~\cite{Bulgac-QMC-2008} & $0$&$152(7)$~\cite{Burovski-QMC-2006}\\
\hline
$\mu(T_c)$ &  $0$&$365$~\cite{PPS-10}  &  $0$&$336$~\cite{PPS-10}  & $0$&$464$~\cite{PPS-10}  &  $0$&$394$~ \cite{Zwerger-07} &     $0$&$43(1)$~\cite{Bulgac-QMC-2008} & $0$&$493(14)$~\cite{Burovski-QMC-2006}\\
\hline
$E(T_c)/N$ &  $0$&$296$~\cite{PPS-10}  &  $0$&$278$~\cite{PPS-10}  & $0$&$337$~\cite{PPS-10}  &  $0$&$304$~ \cite{Zwerger-07} &     $0$&$270(6)$~\cite{Bulgac-QMC-2008} & $0$&$31(1)$~\cite{Burovski-QMC-2006}\\
\hline
\end{tabular}
\caption{Thermodynamic quantities obtained theoretically for a homogeneous Fermi system at unitarity by alternative pairing-fluctuation 
              approaches and QMC calculations (relevant references are specified). 
              Energies are in units of $E_{F}$ and temperatures of $T_{F}$.  
              Here, the dimensionless parameter $\beta$ results from the relation $1 + \beta = E(T=0)/E_{\mathrm{ni}}(T=0)$ that holds
             at unitarity, where $E_{\mathrm{ni}}$ is the energy of the corresponding non-interacting system.}
\label{tabUL_Th}
\end{center}
\end{table}

Alternative theoretical approaches yield different values of the critical temperature $T_{c}$, as shown in Table \ref{tabUL_Th}.
These values can be compared with the corresponding ones that are extracted from experiments, as reported in Table \ref{tabUL_Exp}.
Although for this quantity the self-consistent t-matrix approach seems to perform better than the non-self-consistent one(s),
one should be aware of the fact that additional corrections to the pair propagator $\Gamma^{0}$, like those introduced in Ref.\cite{GMB-61} in the weak-coupling (BCS) limit to represent the medium polarization and shown to have a sizable effect on $T_{c}$ in that limit, might still act to reduce somewhat further the value of $T_{c}$ even at unitarity.
Definite comparison with the experimental values of $T_{c}$ reported in Table \ref{tabUL_Exp} should then await for a proper inclusion of
these additional corrections.
To elicit a more quantitative comparison among theoretical and experimental thermodynamic quantities, Tables \ref{tabUL_Th} and \ref{tabUL_Exp} list, in addition, the values of $\mu$ and $E/N$ that are available both at zero temperature and $T_{c}$.

\begin{table}
\begin{center}
\small
\begin{tabular}[c]{@{} c@{} |r@{.}l@{} |r@{.}l@{} |r@{.}l@{} |r@{.}l@{} |r@{.}l@{} |r@{.}l@{}|}
& \multicolumn{2}{c|}{Exp. \cite{Thomas-2009}} & \multicolumn{2}{c|}{Exp. \cite{Grimm-2004}} & \multicolumn{2}{c|}{Exp. \cite{Jin-2006}} & \multicolumn{2}{c|}{Exp. \cite{Salomon-2009}} & \multicolumn{2}{c|}{Exp. \cite{Hulet-2006}} & \multicolumn{2}{c|}{Exp. \cite{Ketterle-2008}}\\
\hline
$\beta$    & $-0$&$62(2)$          & $-0$&$68_{-0.10}^{+0.13}$ & $-0$&$54_{-0.12}^{+0.05}$ & $-0$&$58(1)$          & $-0$&$54(5)$          & \multicolumn{2}{c|}{}\\
\hline
$\mu(T=0)$ &  $0$&$38(2)$          &  $0$&$32_{-0.10}^{+0.13}$ &  $0$&$46_{-0.12}^{+0.05}$ &  $0$&$42(1)$          &  $0$&$46(5)$          & \multicolumn{2}{c|}{}\\
\hline
$T_c$      & \multicolumn{2}{c|}{} & \multicolumn{2}{c|}{}     & \multicolumn{2}{c|}{}     &  $0$&$157(15)$        & \multicolumn{2}{c|}{} & $\sim0$&$15$\\
\hline
$\mu(T_c)$ & \multicolumn{2}{c|}{} & \multicolumn{2}{c|}{}     & \multicolumn{2}{c|}{}     &  $0$&$49(2)$          & \multicolumn{2}{c|}{} & \multicolumn{2}{c|}{}\\
\hline
\end{tabular}
\caption{Thermodynamic quantities obtained experimentally for a homogeneous Fermi system at unitarity (references are specified). 
             Energies are in units of $E_{F}$ and temperatures of $T_{F}$.  
             The chemical potential at $T=0$ is here obtained via the relation $\mu(T=0)/E_{F} = 1+\beta$ that holds at unitarity,
             the parameter $\beta$ being directly measured. [Experimental data for $E(T_{c})/N$ are not available for comparison with the 
             theoretical values of Table \ref{tabUL_Th}.]}
\label{tabUL_Exp}
\end{center}
\end{table}

The sizable effects that pairing fluctuations have on the thermodynamic quantities of Fig.\ref{figure9} over and above the free-Fermi gas 
behavior can be appreciated by sketching therein the plots of $\mu$ and $E/N$ for the non-interacting Fermi gas (recall, in particular, that
$\mu_{\mathrm{ni}}(T÷=÷0)/E_{F}÷=÷1$, $E_{\mathrm{ni}}(T÷=÷0)/(NE_{F})÷=÷0.6$, $\mu_{\mathrm{ni}}(T÷=÷0.6T_{F})/E_{F}÷=÷0.625$, and
$E_{\mathrm{ni}}(T÷=÷0.6T_{F})/(NE_{F})÷=÷1.15$).
The effects of pairing fluctuations in these quantities are thus seen to extend over a wide temperature range up to several times $T_{F}$, 
being related to the \emph{high-energy scale} $\Delta_{\infty}$ introduced in Ref.\cite{PPS-09} in terms of the trace of the pair propagator 
$\Gamma^{0}$.

There exists, however, an additional energy scale (usually referred to as the \emph{pseudogap}) which is also related to pairing fluctuations but is instead characteristic of the \emph{low-energy} physics about $T_{c}$.
This energy scale is most evident when looking at the properties of the single-particle spectral function, to be considered next.

\subsection{Dynamical properties}
\label{subsec:dyna}

The spectral function $A(\mathbf{k},\omega)$ for single-particle fermionic excitations results after analytic continuation 
of the fermion propagator $G(\mathbf{k},\omega_{n})$ from the Matsubara ($\omega_{n}$) to the real ($\omega$) frequency axis,
via the relation $A(\mathbf{k},\omega) = - \mathrm{Im} \, G^{R}(\mathbf{k},\omega)/\pi$ where $G^{R}(\mathbf{k},\omega)$ is the 
retarded fermion propagator.
Through a related analytic continuation of the fermionic self-energy $\Sigma$, $A(\mathbf{k},\omega)$ can be eventually cast in the form:
\begin{equation}
A(\mathbf{k},\omega) \, = \, - \frac{1}{\pi} \, \frac{\mathrm{Im}\Sigma(\mathbf{k},\omega)}
{[\omega - \xi_{\mathbf{k}} - \mathrm{Re}\Sigma(\mathbf{k},\omega)]^{2} \, + \, 
[\mathrm{Im}\Sigma(\mathbf{k},\omega)]^{2}}                                                              \label{single-particle-spectral-function}
\end{equation}

\noindent
where again $\xi({\mathbf{k}}) = \mathbf{k}^{2}/(2m) - \mu$. For any given wave vector $\mathbf{k}$, the frequency

\begin{figure}[htbp]
\begin{center}
\includegraphics[width=9.0cm,angle=0]{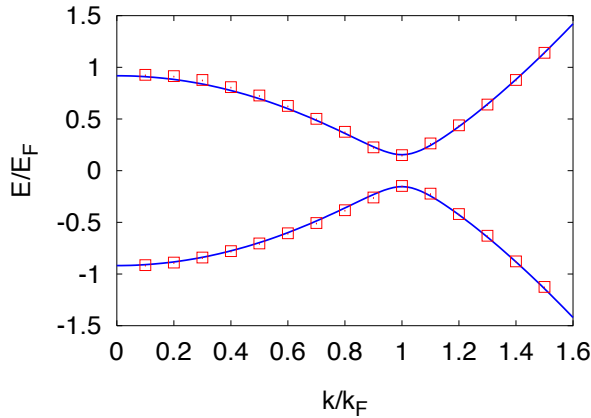}
\caption{Peak positions of the spectral function at $T_{c}$ for negative (lower branch) and positive (upper branch) energies versus 
              the wave vector when $(k_{F} a_{F})^{-1} = - 0.72$. The spectral function is here obtained within the t-matrix approach of 
              Ref.\cite{PPSC-02}. Full lines represent a BCS-like fit. (Adapted from Fig.15 of Ref.\cite{PPSC-02}.)}
\label{figure10}
\end{center}
\end{figure}

\noindent
structure of the real and imaginary parts of $\Sigma(\mathbf{k},\omega)$ 
thus determines the positions and widths of the peaks in $A(\mathbf{k},\omega)$.

The archetype of a pairing-gap behavior for $A(\mathbf{k},\omega)$ is embodied in the two-peak structure of the following expression 
(cf. Eqs.(\ref{BCS-Green-function})): 
\begin{equation}
A(\mathbf{k},\omega) \, = \, u(\mathbf{k})^{2} \, \delta(\omega - E(\mathbf{k})) \, + \, v(\mathbf{k})^{2} \, \delta(\omega + E(\mathbf{k}))
                                                                                                                                      \label{BCS-single-particle-spectral-function}
\end{equation}

\noindent
where $E({\mathbf k})=\sqrt{\xi({\mathbf k})^{2}+\Delta^{2}}$ and $v(\mathbf{k})^{2} = 1 - u(\mathbf{k})^{2} = (1 - \xi({\mathbf{k}})/E(\mathbf{k}))/2$, which holds at the mean-field level in the broken-symmetry phase.
When pairing fluctuations beyond mean field are included \cite{PPSC-02}, a two-peak structure still persists in the normal phase above $T_{c}$, although with broad and asymmetric peaks replacing the delta spikes of Eq.(\ref{BCS-single-particle-spectral-function}) while the total area remains unity.
Even in this case, the positions of the two peaks in the spectral function follow quite closely the BCS-like dispersions
$\pm \sqrt{\xi({\mathbf k})^{2}+\Delta_{\mathrm{pg}}^{2}}$, provided the value $\Delta_{\mathrm{pg}}$ of the pseudogap is inserted
in the place of the BCS gap $\Delta$ of Eq.(\ref{BCS-single-particle-spectral-function}).
An example of this behavior is shown in Fig.\ref{figure10} for weak coupling.

A systematic study of the single-particle spectral function in the normal phase across the BCS-BEC crossover was originally performed in Ref.\cite{PPSC-02} within the t-matrix approach given by Eqs.(\ref{pairing-self-energy})-(\ref{dressed-fermion-propagator}).
Interest in this study was recently revived by the advent of a novel experimental technique for ultracold Fermi gases \cite{Jin-2008}, whereby the wave vector of photo-excited atoms is resolved in radio-frequency spectra taken at different couplings and temperatures.
One should mention in this context the comparison made in Ref.\cite{Levin-09} between theoretical results obtained by the t-matrix and $G$-$G_{0}$ approaches, as well as the calculation performed in Ref.\cite{Zwerger-09} within the fully self-consistent t-matrix approach.

\begin{figure}[htbp]
\begin{center}
\includegraphics[width=9.0cm,angle=0]{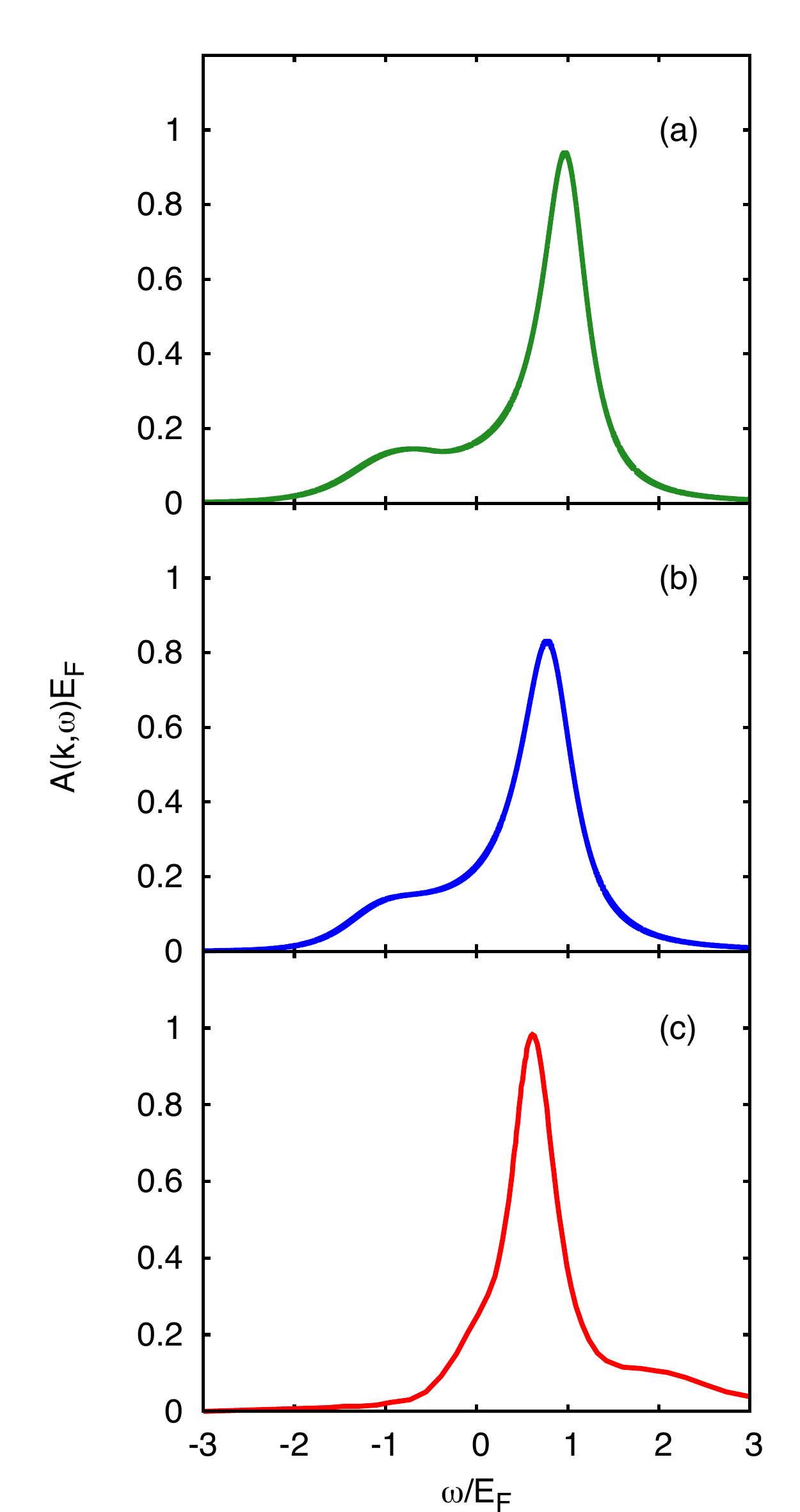}
\caption{Single-particle spectral function at unitarity when $T = T_{c}$ and $k = k_{F}$, as obtained by:
              (a) The t-matrix approach of Ref.\cite{PPSC-02}; (b) The Popov approach of Ref.\cite{PPS-10}; 
              (c) The sc t-matrix approach of Ref.\cite{Zwerger-09}. (The plot of panel (c) has been extracted from
              Fig.4 of Ref.\cite{Zwerger-09}.)}
\label{figure11}
\end{center}
\end{figure} 

Similarly to what was done in subsection~\ref{subsec:thermo} for thermodynamic properties, here we compare the results for $A(\mathbf{k},\omega)$ obtained alternatively by the t-matrix, the Popov, and the fully self-consistent (sc) t-matrix approaches (while referring to 
Ref.\cite{PPS-10} for a more complete analysis of this comparison).
We then show in Fig.\ref{figure11} the results obtained for $A(\mathbf{k},\omega)$ at unitarity and $k = k_{F}$ by the three approaches,
at the respective values of the critical temperature.
[Analytic continuation from Matsubara to real frequencies has been performed in panel (a) by the direct substitution 
$i \omega_{n} \rightarrow \omega + i \eta$, in panel (b) by the Pad\'e approximants, and in panel (c) by the maximum-entropy method.]
Note how the two-peak structure that is evident in panel (a) remains noticeable in panel (b), but has essentially disappeared in panel (c).
This is consistent with a general understanding \cite{Verdozzi-2009} that non-self-consistent calculations favor pseudogap
behavior while self-consistent calculations tend to suppress it.
As the other side of the medal, one would tend to attribute \cite{Zwerger-09} to self-consistent calculations a more precise description of thermodynamic properties with respect to non-self-consistent approaches.

\begin{figure}[htbp]
\begin{center}
\includegraphics[width=10.0cm,angle=0]{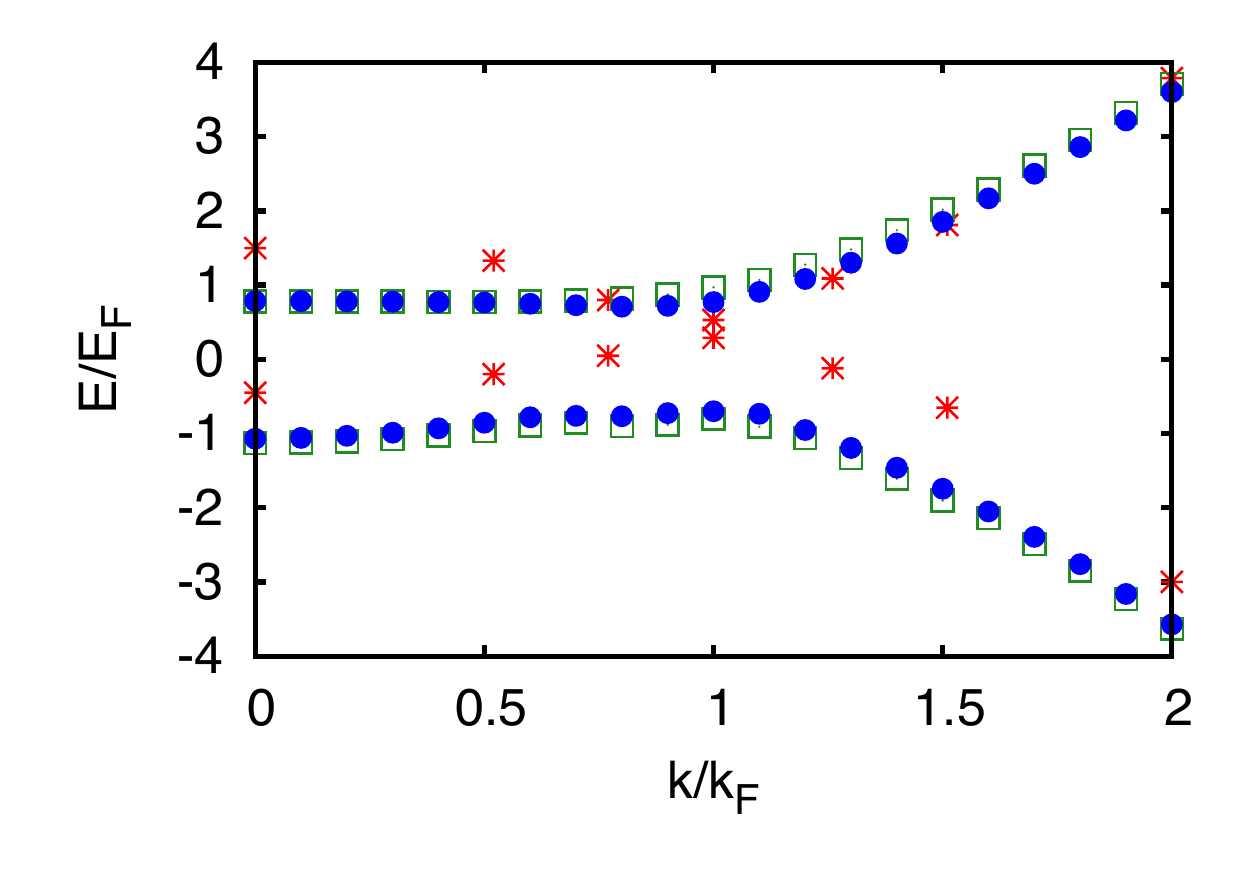}
\caption{Dispersion of the two peaks of the spectral function at unitarity and $T = T_{c}$ obtained by the t-matrix approach 
             (empty squares), the Popov approach (full dots), and the sc t-matrix approach (stars). 
             (Stars have been extracted from the curves in Fig.4 of Ref.\cite{Zwerger-09}.)}
\label{figure12}
\end{center}
\end{figure}

In this respect, it should be noticed that the Popov appproach, while improving considerably on the thermodynamic description with respect to the t-matrix approach as discussed in subsection~\ref{subsec:thermo}, preserves also an evident pseudogap behavior in the single-particle spectral function.
This is made evident in Fig.\ref{figure12}, where the dispersion of the two peaks of $A(\mathbf{k},\omega)$ is reported at unitarity 
and $T = T_{c}$ for the t-matrix, the Popov, and the sc t-matrix approaches.
(Close to $k_{F}$, where the two peaks in $A(\mathbf{k},\omega)$ are broad and overlap each other, the position of the less intense 
peak was determined by subtracting from $A(\mathbf{k},\omega)$ the profile of the most intense peak.)
The value of the pseudogap, identified by (half) the minimum energy separation between the upper and lower branches, remains essentially unmodified when adding the Popov on top of t-matrix fluctuations, but it closes up when full self-consistency is included.

Stringent comparison with both experimental data and QMC calculations will eventually decide what version of pairing-fluctuation theories is able to provide the closest agreement for thermodynamic as well as for dynamical quantities, in systems like ultracold Fermi gases where only the mutual attractive interaction can be responsible for their physical behavior. 

\section{Concluding remarks}
\label{sec:conclusions}

A gas of ultracold Fermi atoms, whose mutual interaction is governed by a (broad) Fano-Feshbach resonance, represents a physical 
system of fermions containing only pairing degrees of freedom. 
This feature naturally conveys their theoretical description in terms of ``pairing fluctuations'' of several kinds in the particle-particle channel, which extend characteristic two-body processes to a finite-density situation.
The difficulty here is that, at finite density, the relevant processes of the pairing type can be unambiguously identified only in the weak-coupling (BCS) regime where a fermion description is appropriate and in the strong-coupling (BEC) regime where a description 
in terms of composite bosons holds, because in both regimes the presence of the small parameter $k_{F} |a_{F}|$ guides the selection
of the diagrammatic contributions for dilute systems.
In additions, in these regimes useful analytic approximations can be quite generally derived from these diagrammatic contributions, 
which help considerably one's physical intuition in picturing the involved processes.
This kind of physical intuition is hard to emerge from more numerically oriented approaches (like QMC calculations) or more abstract approaches (like the renormalization group methods \cite{Wetterich-2009}), thus making diagrammatic approaches to the BCS-BEC crossover more appealing in this respect. 
In principle, diagrammatic approaches may not be controlled in the unitary region, which is intermediate between the BCS and 
BEC regimes and where the diluteness condition does not apply owing to the divergence of $|a_{F}|$.
Nevertheless, the good control of the approximations which can be achieved \emph{separately} in the BCS and BEC regimes \emph{and} 
the limited extension of the unitary region ($- 1 \lapprox (k_{F} a_{F})^{-1} \lapprox + 1$) enable one to formulate a single fermionic theory that bridges across the BCS and BEC regimes and is able to furnish a good description of the unitary region for most practical purposes.
This is the spirit with which diagrammatic pairing-fluctuation approaches to the BCS-BEC crossover have been formulated and applied to a variety of problems, both in the normal and superfluid phases.

It is finally relevant to mention that the interest in the physics brought about by consideration of pairing fluctuations is not limited to a system of ultracold Fermi atoms.
In particular, the issue of the possible occurrence of a pseudogap in single-particle excitations is of considerable interest both in condensed 
matter \cite{Damascelli-2002} and nuclear physics \cite{Bozek-1999}.
This gives to ultracold Fermi gases the role of prototype systems, in which issues of general interest can be conveniently addressed 
by exploiting the unprecedented flexibility that they provide in the control of their physical parameters.
\vspace{0.8cm}

\noindent
{\bf Acknowledgements}

\noindent
The author is indebted to F. Palestini for a critical reading of the manuscript.

\vspace{2.0cm}          

\end{document}